\documentstyle[12pt]{article}
\newcommand{\beq}{\begin{equation}}
\newcommand{\eeq}{\end{equation}}
\newcommand{\ba}{\begin{eqnarray}}
\newcommand{\ea}{\end{eqnarray}}

\newcommand{\dsl}
  {\kern.06em\hbox{\raise.15ex\hbox{$/$}\kern-.56em\hbox{$\partial$}}}

\newcommand{\eeqarr}{\end{eqnarray}}
\newcommand{\ZZ}{{\rm \kern 0.275em Z \kern -0.92em Z}\;}
\begin{document}
\begin{titlepage}
\begin{center}
{\Huge Energy and the AdS/CFT Correspondence}
\\
\vspace*{0.5cm}
{\large Pablo Minces$^{\dag,}$\footnote{pablo@fma.if.usp.br} and
Victor O. Rivelles$^{\ddag,}$\footnote{rivelles@fma.if.usp.br}}
\\
\vspace*{0.2cm}
$^{\dag}$Instituto de F\'{\i}sica Te\'orica, Universidade Estadual
Paulista\\
Rua Pamplona 145, 01405-900, S\~ao Paulo, SP, Brasil\\
\vspace*{0.2cm}
$^{\ddag}$Instituto de F\'{\i}sica, Universidade de S\~ao Paulo\\
Caixa Postal 66.318, 05315-970, S\~ao Paulo, SP, Brasil
\vspace*{0.5cm}

\end{center}
\begin{abstract}
We consider a scalar field theory on AdS in both minimally and
non-minimally coupled cases. We show that there exist constraints 
which arise in
the quantization of the scalar field theory on AdS which cannot
be reproduced through the usual AdS/CFT prescription. We argue that the 
usual energy, defined through the stress-energy
tensor, is not the natural one to be considered in the context of the
AdS/CFT correspondence. We analyze a new definition of the energy
which makes use of the Noether current corresponding to time
displacements in global coordinates. We compute the new energy for
Dirichlet, Neumann and mixed boundary conditions on the scalar
field and for both the minimally and non-minimally coupled cases. Then, we
perform the quantization of the scalar field theory on AdS showing 
that, for `regular'
and `irregular' modes, the new energy is conserved, positive and
finite. We show that the quantization gives rise, in a natural way,
to a generalized AdS/CFT prescription which maps to the boundary all
the information contained in the bulk. In
particular, we show that the divergent local terms of the on-shell
action contain information about the Legendre transformed generating
functional, and that the new constraints for which the irregular modes
propagate in the bulk are the same constraints for which such divergent
local terms cancel out. In this situation, the addition of counterterms is
not required. We also show that there exist particular cases for which the
unitarity bound is reached, and the conformal dimension becomes
independent of the effective mass. This phenomenon has no bulk
counterpart.

\end{abstract}

\vskip 0.5cm

\begin{flushleft}
\end{flushleft}
\end{titlepage}

\section{Introduction}

Since the proposal of Maldacena conjecturing the existence of a duality
between 
Type IIB supergravity theory on Anti-de Sitter (AdS) space and the large
N limit of a conformal ${\cal N}=4$ Super Yang-Mills theory
living on the boundary of the AdS space \cite{maldacena}, a large
amount of 
work has been devoted to understand all the implications of the
so-called AdS/CFT correspondence. For the d+1 dimensional AdS space
($AdS_{d+1}$), the explicit prescription for mapping
one theory into the other has been given in \cite{witten}\cite{gubser}, 
and
it takes the form

\begin{equation}
Z _{AdS}[\phi _{0}] = \int _{\phi _{0}}{\cal D}\phi\
\exp\left(-I[\phi]\right)\equiv Z _{CFT}[\phi _{0}] = \left<\exp\left(\int
_{\partial\Omega}d^{d}x{\cal O}\phi _{0}\right)\right>\; ,
\label{9003}
\end{equation}
where $\phi_{o}$ is the boundary value of the bulk field $\phi$ which
couples to the boundary conformal field theory (CFT) operator ${\cal
O}$. In this prescription, the path integral
over the bulk fields on the l.h.s is computed under the restriction
that the field $\phi$ satisfies a Dirichlet boundary condition on the
boundary of AdS. The above Dirichlet prescription has been
successfully
applied to
the cases of the scalar field
\cite{witten}\cite{freedman3}\cite{viswa1}\cite{kraus}\cite{our2},
the
spinor
field
\cite{sfetsos}\cite{viswa2}\cite{henneaux2}\cite{frolov1}, the vector
field
\cite{witten}\cite{freedman3}\cite{viswa2}\cite{our}\cite{riv}, the
Rarita-Schwinger field \cite{volovich}\cite{koshelev}\cite{viswa4},
the graviton
field \cite{liu}\cite{viswa5}, the massive symmetric tensor field
\cite{polishchuk}
and the antisymmetric $p$-form field \cite{frolov}\cite{l'yi}. In
all cases, the boundary CFT two-point and higher 
order functions obtained through the mapping Eq.(\ref{9003}) were shown to
have the form prescribed by conformal invariance, and the conformal
dimensions of the boundary operators corresponding to all the fields
listed above were established. We remark that, in all
references above but \cite{our2}, only Dirichlet boundary conditions
were employed. Further refinements include the addition of
local counterterms to
the action
\cite{frolov1}\cite{skenderis}\cite{witten2}\cite{nojiri}\cite{odintsov2}\cite{kraus2}\cite{johnson}\cite{viswa6}\cite{petkou}\cite{haro}
which cancel the divergences arising when going to the
boundary. We also mention the work on
holographic renormalization group flows from deformations on AdS which
extends the AdS/CFT correspondence to non-conformal field theories
\cite{akhmedov}\cite{alvarez}\cite{freedman4}\cite{verlinde}. 

In this work, we concentrate on the formulation of the scalar
field theory on AdS spaces in both minimally and non-minimally coupled 
cases. It is known \cite{freedman}\cite{freedman8} that the usual
quantization of the scalar field theory on AdS spaces in global
coordinates\footnote {For recent work on 
quantization in Poincar\'e coordinates, see
\cite{boschi1}\cite{boschi2}\cite{boschi3}} involves two different kinds
of
normalizable modes, namely the `regular' and `irregular' ones. Since
there are two possible consistent quantizations of the scalar field
on AdS, we expect that for each one of them there exists a
corresponding boundary CFT. But, as pointed out in \cite{witten2}, the 
Dirichlet prescription Eq.(\ref{9003}) accounts only for the CFT that
corresponds to the situation in which regular modes
propagate in the bulk. In order to also account for the missing CFT, the
proposal in \cite{witten2} is that its generating functional can
be
obtained by Legendre transforming the original one which corresponds to
regular modes. It has
been shown in \cite{witten2} that, in fact, this procedure gives rise 
to a boundary CFT whose conformal
dimension is the one expected for irregular modes propagating
in the bulk.

Then, at first sight, it seems that the usual quantization of the
scalar field theory on AdS along the lines of
\cite{freedman}\cite{freedman8}\cite{mezincescu}, together with the
AdS/CFT formulation which makes use of the Dirichlet prescription
Eq.(\ref{9003}) and the Legendre transform prescription
in \cite{witten2}, gives rise to a consistent formulation of the
scalar field theory on AdS spaces and in the AdS/CFT
correspondence. However, we will show in this
article that the usual quantization in global coordinates, as developed in
\cite{freedman}\cite{freedman8}\cite{mezincescu}, imposes constraints on
the mass and the coupling coefficient
of the field to the metric which cannot be mapped to the boundary through
the usual
Dirichlet and Legendre transform prescriptions. It is a serious drawback, 
because the very first thing that we require on
any AdS/CFT prescription is that it must be able to map to the boundary
all the
information contained in the AdS bulk. Then, in order to
remove these difficulties, we will argue that the usual energy, which is
constructed through the `improved' stress-energy tensor, as in
\cite{freedman}\cite{freedman8}\cite{mezincescu}, is not the natural one
to
be considered in the AdS/CFT correspondence context. This will require
a new definition of energy, associated to the Noether current
corresponding to time displacements, which is the natural one to be
considered in the context of the AdS/CFT correspondence. As a result, we 
will perform, for both the minimally and non-minimally coupled cases, 
a new consistent quantization of the scalar field theory on AdS in global
coordinates. We will show that this definition of energy is sensitive
to the boundary conditions. Thus, we will compute it for all
possible boundary conditions on the scalar field, namely Dirichlet,
Neumann,
and a combination of both of them which we call mixed boundary
condition, but that sometimes is also called as Robin or third boundary
condition. We will show that the energy is conserved, positive and finite
for regular and
irregular modes arising for some constraints on the mass and the
coefficients of mixed boundary conditions thus leading to a consistent
quantization of the scalar field theory on the AdS bulk in global
coordinates. Besides, in this new picture the Breitenlohner-Freedman
bound \cite{freedman}\cite{freedman8}\cite{mezincescu} still holds, as
expected.

Then, in the AdS/CFT context, we will look for
generalizations of the usual Dirichlet and Legendre transform
prescriptions for which the boundary CFT's contain all the
information about the new constraints just mentioned and the regular and
irregular modes propagating in the bulk. We will show that the
incorporation of boundary conditions leads, in a natural way, to a
generalized AdS/CFT prescription in
which conformal operators couple at the boundary to sources which
depend on the selected boundary condition.

We will also consider a generalized prescription for the Legendre
transform, in which we transform the whole action at the boundary, rather
than only the leading non-local term, as in the usual prescription. This
generalized prescription makes use, in fact, of the standard form of
the Legendre transformation. At first sight, it could seem that such a
generalization of the usual prescription introduces no new
results. However, this is not the case,
because, as we will show, the divergent local terms of the on-shell
action contain information about the Legendre transformed
generating functional, and then, the operations of expanding in powers of
the
distance to the boundary and then selecting the generating functional, and
of performing the Legendre transformation, are not commuting
operations. As a
consequence, we will show that the generalized prescription
gives rise to results which, in general, are different from those arising
from the usual Legendre transform prescription. A key result will be that
the new results arising from the generalized AdS/CFT prescription are in 
whole agreement with the ones that we find on performing the 
quantization in the bulk. In addition, this generalized
Legendre transform prescription will let us to remove some difficulties 
that appear in the usual prescription.

In general, we will show that our formulation gives rise to boundary CFT's
which contain
all the information about the quantization of the scalar
field theory on AdS in global coordinates, and, in particular, we will
also show that the constraints for which the irregular modes
propagate in the bulk are the same constraints for which the
divergent local terms of the on-shell action cancel out. In this
situation, the addition of counterterms is not required. We will
also show that there exists one particular case which
has no bulk counterpart, namely the one in which the conformal dimension
reaches the unitarity bound and becomes independent of the effective mass.

The paper is organized as follows. In Section 2, we motivate the
introduction of a new formalism for the scalar field theory on AdS and in
the AdS/CFT correspondence. We consider some difficulties regarding the
usual formulation, and provide some clues for a new formalism in which
such difficulties can be removed. In Section 3, we compute the
stationary
actions corresponding to Dirichlet, Neumann and mixed boundary conditions
on the scalar field in both the minimally and non-minimally coupled cases,
and then, for each action, we construct the Noether current corresponding
to time displacements. We show that it is sensitive to the
addition of boundary terms to the action, and that it is conserved, 
in the sense that its covariant divergence vanishes. Then, we
compute the energy associated to the Noether current, and find the
constraints for which it is conserved,
positive and finite. In this way, we can consistently quantize the scalar 
field theory on AdS. In particular, we
compute the new constraints for which the irregular modes propagate in
the bulk. In Section 4, we consider the generalized AdS/CFT prescription,
and show that it leads to results which are in complete agreement
with those obtained
by performing the quantization in global coordinates. In
particular, we show that the new constraints for which irregular modes
propagate in the bulk are reproduced, in a natural way, by the
generalized AdS/CFT prescription. This is a remarkable non-trivial
result which can be considered as a strong evidence in support of our
proposed formalism. We will also show that the divergent local terms of
the on-shell action contain information about the Legendre transformed
generating functionals, and that the new constraints for which the
irregular modes propagate in the bulk are the same for which
such divergent local terms cancel out, so that no counterterms are 
required. Finally, Section 5 presents our
conclusions. Appendix A contains some useful identities for the
hypergeometric functions, and Appendix B
presents the main results obtained from the generalized AdS/CFT
prescription.

\section{Motivations}

In order to set up the notation and motivate the subsequent development,
we begin this section by reviewing the main results of the usual
formulation of scalar field theory on AdS spaces and in the AdS/CFT
correspondence. Most of the results presented in this section will
be used throughout this paper. 

A consistent quantization of a massive real
scalar field theory on $AdS_{4}$ was found in
\cite{freedman}\cite{freedman8} (see also
\cite{avis} for earlier results) and then extended to $AdS_{d+1}$ in 
\cite{mezincescu}. It will be useful to begin by concentrating on
the minimally coupled case. Consider the usual action in $AdS_{d+1}$

\beq
-\frac{1}{2}\;\int d^{d+1} x \;\sqrt{g}\;
\left(
g^{\mu\nu}\partial_{\mu}\phi\;\partial_{\nu}\phi+
m^{2}\phi^{2}\right) \; ,
\label{9007}
\eeq
where $m$ is the mass of the scalar field, and $g_{\mu\nu}$ is a metric 
of `mostly
plus'
signature describing $AdS_{d+1}$. In the formulation of
\cite{freedman}\cite{freedman8}\cite{avis}, the 
solution of the classical equation of motion of the scalar field
$\left(\nabla^{2}-m^{2}\right)\phi =0$ is expanded in modes

\beq
\phi=\sum_{n}\left[ a_{n}\phi_{n}
+a_{n}^{*}\phi_{n}^{*}\right]\; ,
\label{9008}
\eeq
for which the following scalar product is defined

\beq
\left(\phi_{n},\phi_{n'}\right)=
i\int d^{d}x \sqrt{g} \; g^{0\nu}
\left[
\phi_{n}\partial_{\nu}\phi^{*}_{n'}-
\partial_{\nu}\phi_{n}\phi^{*}_{n'}\right]\; .
\label{9005}
\eeq
Here, the $a_{n}$'s are arbitrary complex coefficients, and $n$ labels the
modes. The above scalar product is the `conserved' charge
corresponding to
the conserved current (in the sense that its covariant divergence
vanishes)

\beq
H^{\mu}=ig^{\mu\nu}\left[
\phi_{n}\partial_{\nu}\phi^{*}_{n'}-
\partial_{\nu}\phi_{n}\phi^{*}_{n'}\right]\; .
\label{9004}
\eeq
The problem which arises when quantizing the scalar field theory on AdS is
that this space does not admit a Cauchy surface, and a massless
particle reaches the spatial infinity in a finite time. The solution
proposed in \cite{freedman}\cite{freedman8}\cite{avis} for performing a
consistent quantization was to compactify the space and impose
appropriate
boundary conditions on the solution of the classical equation of motion,
such as the scalar product Eq.(\ref{9005}) be actually conserved. This 
means to require that there is no flux of the current
Eq.(\ref{9004}) through the boundary at spatial infinite. This imposes
further conditions on the solution of the equation of motion, and the
result
in \cite{freedman}\cite{freedman8}\cite{mezincescu} is 
that there exist two different asymptotic behaviors of the scalar
field such that the flux at the boundary vanishes. They are called
`regular'
and `irregular' modes, and they behave close to the boundary as

\beq
\phi_{R}\sim{\hat\epsilon}^{\Delta^{(0)}_{+}}\; ,\quad \phi_{I}\sim
{\hat\epsilon}^{\Delta^{(0)}_{-}}\; ,
\label{9006}
\eeq
where $\phi_{R}$, $\phi_{I}$ correspond to regular and irregular
modes respectively, ${\hat\epsilon}$ is a measure of the distance to the
boundary which is considered to be small, and

\beq
\Delta^{(0)}_{\pm}=\frac{d}{2}\pm\nu^{(0)}\; ,
\label{9010}
\eeq

\beq
\nu^{(0)}=\sqrt{\frac{d^2}{4}\;+\;m^{2}}\; .
\label{9011}
\eeq
One important result in \cite{freedman}\cite{freedman8} is
that, whereas for regular modes the flux
of the current Eq.(\ref{9004}) vanishes for any $\nu^{(0)}\geq 0$,
for the irregular ones it vanishes only for

\beq
0\leq \nu^{(0)} < 1\; .
\label{9010'''}
\eeq
Since regular and irregular modes are normalizable, the scalar field can 
be consistently quantized. Note that for $\nu^{(0)} >1$ there is only one
possible quantization,
namely the one corresponding to regular modes. On the other hand, for
$0\leq\nu^{(0)} <1$ both regular and irregular modes are normalizable, and
then we have two possible
consistent quantizations. In spite of the fact that the regular solution
has been much more studied than the irregular one, both of them exist and
lead to consistent formulations. It is also important to point out that
stability imposes the condition
that $\nu^{(0)}$ is a real number 
\cite{freedman}\cite{freedman8}\cite{mezincescu}. Such a condition can
be written as

\beq
m^{2}\geq -\frac{d^{2}}{4}\; ,
\label{9011'}
\eeq
which is known as the Breitenlohner-Freedman bound.

The analysis of regular and irregular modes in the context of the
AdS/CFT correspondence was performed in \cite{kraus}, where it has been
suggested that there are two kinds of modes which are relevant to the
formulation of the scalar field theory on AdS, namely the quantum,
fluctuating modes which propagate in the bulk, and the
classical,
non-fluctuating backgrounds which couple to the boundary conformal
operators
through the AdS/CFT prescription Eq.(\ref{9003}). For $\nu^{(0)} >1$, only
regular modes are normalizable, and thus they are the only kind of
modes which propagate in the bulk. In this
situation, the irregular modes are the classical, non-fluctuating 
backgrounds which couple to the boundary conformal
operator, that happens to have conformal dimension
$\Delta^{(0)}_{+}$. This
result has been successfully reproduced through the Dirichlet
prescription Eq.(\ref{9003}) \cite{witten}\cite{freedman3}\cite{viswa1}.

However, a new feature arises for $0\leq\nu^{(0)} <1$. In this situation,
both
regular and irregular modes are normalizable, and thus we have two
possible consistent quantizations. Since the AdS/CFT correspondence states
that any
quantum field theory on AdS is equivalent to a conformal theory on the
boundary, we expect that we should have two different boundary CFT's. One
of them corresponds to the situation in which we choose the regular
modes to describe the quantum fluctuations propagating in the bulk. In
this case, we expect to have a boundary
conformal operator of conformal dimension $\Delta^{(0)}_{+}$, and this
CFT can be reproduced through the Dirichlet prescription Eq.(\ref{9003}),
as
it happens for $\nu^{(0)} >1$. The problem arises when we choose the
irregular modes to describe the quantum fluctuations propagating in the 
bulk. In this situation, we
expect the boundary conformal operator to have conformal dimension
$\Delta^{(0)}_{-}$. However, as pointed out in \cite{witten2}, such an
operator
cannot be reproduced through the Dirichlet prescription
Eq.(\ref{9003}). It is also important to note that the conformal
dimension
$\Delta^{(0)}_{+}$ is bounded from below by $d/2$, whereas the
constraint Eq.(\ref{9010'''}) makes the conformal dimension
$\Delta^{(0)}_{-}$ to be
bounded from below by $(d-2)/2$, which is precisely the unitarity bound
for scalar operators in d dimensional field theory. The bound $d/2$
is more stringent than the unitarity bound $(d-2)/2$, and in fact there
exist particular examples for which the conformal dimension
$\Delta^{(0)}_{-}$
is needed (for details, see \cite{witten2} and references therein). 

From all considerations above, we conclude that, in order to be able to
reproduce not only the conformal dimension $\Delta^{(0)}_{+}$ but also the
missing conformal dimension $\Delta^{(0)}_{-}$, some generalization
of the Dirichlet prescription Eq.(\ref{9003}) must be introduced. In
doing so, the proposal in \cite{witten2} is that the generating functional
in the theory with conformal dimension $\Delta^{(0)}_{-}$ is a Legendre
transform of the generating functional in the theory with conformal
dimension $\Delta^{(0)}_{+}$ (see also \cite{dobrev} for previous results
arising from group-theoretic analysis). It is known
\cite{witten}\cite{freedman3}\cite{viswa1} that such a generating
functional is written in momentum space as

\beq
S[\phi_{0}]\; =\;\frac{\Gamma
(1-\nu)}{\Gamma(\nu)}\;\int\frac{d^{d}k}{(2\pi)^{d}}\;
\phi_{0}\left(\vec{k}\right)\phi_{0}\left(-\vec{k}\right)\;
\left(\frac{k}{2}\right)^{2\nu}\; ,
\label{9012}
\eeq
where $k=\mid\vec{k}\mid$. The prescription in \cite{witten2} is to
carry out the Legendre transformation by setting

\beq
{\tilde S}[\phi_{0},{\tilde \phi}_{0}]\; = S[\phi_{0}]\; +\; 
\alpha\int\frac{d^{d}k}{(2\pi)^{d}}\;
\phi_{0}\left(\vec{k}\right){\tilde \phi}_{0}\left(-\vec{k}\right)\; ,
\label{9013}
\eeq
where the coefficient $\alpha$ has been chosen in \cite{witten2} to be 
$\alpha =2\Delta^{(0)} -d$. The generating functional in the theory with
conformal dimension $\Delta^{(0)}_{-}$ is the minimum of
${\tilde S}[\phi_{0},{\tilde \phi}_{0}]$ with respect to $\phi_{0}$ (for
fixed
${\tilde \phi}_{0}$), and it is given by

\beq
{\tilde S}[{\tilde \phi}_{0}]\; =\;-\frac{\alpha^{2}}{4}\;\frac{\Gamma
(\nu)}{\Gamma(1-\nu)}\;\int\frac{d^{d}k}{(2\pi)^{d}}\;
{\tilde
\phi}_{0}\left(\vec{k}\right){\tilde \phi}_{0}\left(-\vec{k}\right)\;
\left(\frac{k}{2}\right)^{-2\nu}\; .
\label{9014}
\eeq
Integration over the momentum shows that the field ${\tilde
\phi}_{0}$ couples to a boundary conformal operator
${\tilde {\cal O}}$ of conformal dimension $\Delta^{(0)}_{-}$ through the
prescription \cite{witten2}

\beq
\exp\left( -{\tilde S}\right) \equiv \left<\exp\left(\int d^{d}x \;
{\tilde {\cal O}}(\vec{x}) \; {\tilde \phi}_{0}(\vec{x})\right)\right>,
\label{9015}
\eeq
which arises from the prescription Eq.(\ref{9003}) by considering the
transformed field ${\tilde \phi}_{0}$ as the source for the new
boundary conformal operator ${\tilde {\cal O}}$.

Thus, at first sight, it seems that the AdS/CFT prescription
Eqs.(\ref{9003}, \ref{9013}) successfully maps to the boundary all the
information
obtained in \cite{freedman}\cite{freedman8}\cite{mezincescu} on performing
the quantization of the scalar field theory on
the AdS
bulk. However, one of the purposes of this work is to point out that there
remain some open problems which need to be
considered. One of them is the fact that the constraint 
Eq.(\ref{9010'''}) for which the irregular modes are normalizable has not
so far
been reproduced by any AdS/CFT prescription. This is so because the
Legendre transformation Eq.(\ref{9013}) can, in principle, be carried out
for
any
value of $\nu$. Thus, in the usual prescription, the constraint
Eq.(\ref{9010'''}) has to be imposed `by hand'. This is a serious
drawback, because the very
first thing that we require on any AdS/CFT
prescription is that it must be able to map to the boundary all the
information 
contained in the bulk. In this particular case, we would
expect that the AdS/CFT prescription could reproduce the constraint
Eq.(\ref{9010'''}) in a natural way, just as the analysis along the lines
of \cite{freedman}\cite{freedman8} does in the bulk.

There is still another problem regarding the usual
Dirichlet and Legendre transform prescriptions
Eqs.(\ref{9003}, \ref{9013}), related to the existence of a further
constraint which they cannot reproduce, and that needs to be imposed `by
hand' too. Such a constraint
arises on considering the role of the energy in the usual quantization of
the scalar field theory on AdS, along the lines of
\cite{freedman}\cite{freedman8}\cite{mezincescu}. In order to make this
clear, we begin by considering, instead of Eq.(\ref{9007}), a new action
in which the scalar field is now non-minimally coupled to the $AdS_{d+1}$
metric

\beq
I_{matter}=-\frac{1}{2}\;\int d^{d+1} x \;\sqrt{g}\;
\left[
g^{\mu\nu}\partial_{\mu}\phi\;\partial_{\nu}\phi\; +
\;\left( m^{2}+\;\varrho R\right)\phi^{2}\right] \; ,
\label{9016}
\eeq
where $R$ is the Ricci scalar corresponding to $AdS_{d+1}$ (note that 
it is a constant), and $\varrho$
is an arbitrary coupling coefficient. In this situation, the non-minimal
coupling introduces a new effective mass

\beq
M^{2}(\varrho)=m^{2}+\;\varrho R\; ,
\label{9016'}
\eeq
and now the regular and irregular modes for which the current
Eq.(\ref{9004}) is conserved behave, close to the boundary, as follows

\beq
\phi_{R}\sim{\hat\epsilon}^{\Delta_{+}(\varrho)}\; ,\quad
\phi_{I}\sim {\hat\epsilon}^{\Delta_{-}(\varrho)}\; ,
\label{9006'}
\eeq
where

\beq
\Delta_{\pm}(\varrho)=\frac{d}{2}\;\pm\;\nu(\varrho)\; ,
\label{9006''}
\eeq
\beq
\nu(\varrho)=\sqrt{\frac{d^2}{4}\;+\;M^{2}(\varrho)}\; .
\label{9006'''}
\eeq
In particular, the coefficients $\Delta_{\pm}^{(0)}$ and $\nu^{(0)}$ in
Eqs.(\ref{9010}, \ref{9011}) are just the coefficients 
$\Delta_{\pm}(0)$ and $\nu(0)$ respectively, and the constraint
Eq.(\ref{9010'''}) for irregular modes is generalized as follows

\beq
0\leq \nu(\varrho) < 1\; .
\label{12000}
\eeq
Note that the Breitenlohner-Freedman bound now reads

\beq
M^{2}(\varrho)\geq -\frac{d^{2}}{4}\; ,
\label{9011''}
\eeq
instead of Eq.(\ref{9011'}).

Under an infinitesimal variation of the metric 

\beq
g_{\mu\nu}\rightarrow g_{\mu\nu} + \delta g_{\mu\nu}\; ,
\label{9017}
\eeq
the action Eq.(\ref{9016}) transforms as

\beq
\delta_{g} I_{matter}\; =\; -\;\frac{1}{2}\;\int d^{d+1}x
\;\sqrt{g}\;T_{\mu\nu}\;\delta g^{\mu\nu} +\; \mbox{surface terms} ,
\label{9018}
\eeq
where $T_{\mu\nu}$ is the `improved' stress-energy tensor of the scalar
field. It is given by

\beq
T_{\mu\nu}= \partial_{\mu}\phi\;\partial_{\nu}\phi \;-\;
\frac{1}{2}\; g_{\mu\nu}\left[g^{\alpha\beta}\partial_{\alpha}\phi
\;\partial_{\beta}\phi + M^{2}(\varrho)\;\phi^{2}\right]\; +\;\varrho
\; (g_{\mu\nu}\nabla^{2}-\nabla_{\mu}
\nabla_{\nu}+ R_{\mu\nu})\;\phi^{2}\; ,
\label{9019}
\eeq
where $R_{\mu\nu}$ is the Ricci tensor of $AdS_{d+1}$ and
$\nabla_{\mu}$ stands for a covariant derivative. The analysis in
\cite{freedman}\cite{freedman8} is based on a previous formulation in 
\cite{avis}\cite{abbott} and, in particular, it defines the contribution
of
the scalar field to the
energy of the system as the `conserved' charge corresponding to the
conserved current

\beq
J^{\mu}=-T^{\mu}_{\; 0}\; ,
\label{9020}
\eeq
where we have contracted the stress-energy tensor with the Killing vector
corresponding to time displacements in global coordinates. Note that we 
have included
a minus sign because of the `mostly plus' signature of the
metric. For future purposes, it will be useful to call the
`conserved' quantity described above as the `metrical
energy'. In order for the `conserved' metrical energy to be
actually conserved, Breitenlohner and Freedman impose that there is no
flux
of the current Eq.(\ref{9020}) through the boundary (see also
\cite{mezincescu} for the extension
to the d+1 dimensional case), and this
restriction fixes further conditions on
the solution of the equation of motion. The result is
that, whereas for regular modes the metrical energy is actually conserved
for any value of $\varrho$, for irregular modes it is conserved only for
$\varrho$ being a solution of the constraint

\beq
\varrho = \frac{1}{2}\;\frac{\Delta_{-}(\varrho)}
{1+2\Delta_{-}(\varrho)}\; .
\label{9041}
\eeq
This new constraint must be added to the usual one Eq.(\ref{12000}) that
we have found on considering the conservation of the scalar product
Eq.(\ref{9005}) for irregular modes. 

A very important result
in \cite{freedman}\cite{freedman8}\cite{mezincescu} is that for regular
modes and irregular ones satisfying Eqs.(\ref{12000}, 
\ref{9041}), and only for them, the metrical energy is not only
conserved, but also positive and finite. This leads to a consistent
quantization of the scalar field theory on AdS in the non-minimally
coupled case. In particular, for $0\leq \nu<1$ we have again two
possible quantizations corresponding to the situations in
which the modes which propagate in the bulk are chosen to be the
regular or the irregular ones. On the other hand, for $\nu>1$
we have only one
possible consistent quantization, namely the one in which the regular
modes propagate in the bulk.

Now we are ready to explain the second problem regarding the
usual Dirichlet and Legendre transform prescriptions
Eqs.(\ref{9003}, \ref{9013}). Note that in 
such formulation, the
coupling between the scalar field and the metric is understood just as a
mere renormalization of the mass (see Eq.(\ref{9016'})). In particular,
the
Legendre transformation Eq.(\ref{9013}) can be performed for any value of
$\varrho$, and thus, in this formalism, the constraint
Eq.(\ref{9041}) cannot be
reproduced, just as it happens to the constraint Eq.(\ref{12000}) as
pointed out before. Again, we are facing a
serious drawback.

Summarizing, the Legendre transform prescription Eq.(\ref{9013}) is
able to reproduce the missing conformal dimension $\Delta_{-}(\varrho)$,
but it has no information about the constraints on $\nu$ and
$\varrho$ for which irregular modes propagate in the bulk.

Furthermore, there still remain two more problems regarding the usual
Legendre
transform prescription. One of them is that it
does not fix the
coefficient $\alpha$ in Eq.(\ref{9013}). In \cite{witten2}, it has been
chosen to be $\alpha =2\Delta -d$ under the conjecture
that the conjugated field of $\phi_{0}$ is actually $(2\Delta -d){\tilde
\phi}_{0}$ rather than ${\tilde \phi}_{0}$.\footnote{The calculation
in \cite{witten2} involves taking the derivative of the renormalized
on-shell 
action and compute the expectation value of the dual operator to the
field. Then, after having computed the coefficient $\alpha$, it is
included in the Legendre transform prescription.} We are looking
for a prescription which fixes $\alpha$ in a natural way, thus leaving no
coefficient to be chosen `by hand'. This means that, instead of including
$\alpha$ in the Legendre transform prescription, we want the Legendre
transformation itself to fix the precise form of the conjugated field.

And the final problem regarding the usual AdS/CFT prescription
is related to the fact that for integer $\nu$ the
on-shell action also contains logarithmic terms which are not   
present
for non-integer $\nu$. For instance, for $\nu=0$ the generating
functional is written in momentum space as \cite{our2}

\beq
S[\phi_{0}]\; =\;\frac{1}{2}\;\int\frac{d^{d}k}{(2\pi)^{d}}\;
\phi_{0}\left(\vec{k}\right)\phi_{0}\left(-\vec{k}\right)\;
ln\; k\; ,
\label{9030}
\eeq
instead of Eq.(\ref{9012}). The case $\nu=0$ is included in the
interval Eq.(\ref{12000}) for which irregular modes propagate in the 
bulk. In fact, this is the particular case for which the conformal
dimensions $\Delta_{+}$ and $\Delta_{-}$ are
equal. Thus, for
$\nu =0$ we
would expect the generating functional to be self-conjugated. However,  
it is easy to verify that, in this case, the usual Legendre transform
prescription Eq.(\ref{9013}) does not work, due to the presence of  
the logarithmic term in Eq.(\ref{9030}). This indicates that something
is being missed in the usual formulation.

For all these reasons, we will propose a new formulation of the scalar
field theory on AdS and in the AdS/CFT 
correspondence, in both minimally and non-minimally coupled cases, for
which all the difficulties that arise from the usual formulation can be
removed. In order
to do this, we begin by concentrating on the definition of the energy of
the scalar field theory on the AdS bulk. Note that the AdS/CFT
correspondence
is sensitive to the addition of boundary terms to the action. This can be
seen to be true by computing the l.h.s of Eq.(\ref{9003}) for a  
classical field configuration. All that is left is a boundary term. If we 
start with different boundary terms in the action then we obtain different
correlation functions on the r.h.s. But if the AdS/CFT
correspondence is sensitive to the addition of boundary terms, and we
require that it must also contain all the information regarding the
conservation, positivity and finiteness conditions on the energy of the
theory on the
bulk, then it is natural to demand that the energy of the theory on the
bulk has also to be sensitive to the addition of boundary terms to the
action. However, the metrical energy, which is the one that is analyzed in
the usual formulation along
the lines of \cite{freedman}\cite{freedman8}\cite{mezincescu},
does not fulfil this requirement. This is so
because, as it can easily be verified, the addition of a boundary term to
the action Eq.(\ref{9016}) amounts to the addition of only new
surface
terms in Eq.(\ref{9018}), and it means that the stress-energy tensor
Eq.(\ref{9019}) is not sensitive to the addition of boundary
terms to the action.\footnote{One could argue that the surface terms in
Eq.(\ref{9018}) contribute to the stress-energy tensor with terms
containing delta functions. However, such surface terms vanish, because
we always impose proper boundary conditions. This topic will be discussed
in detail in Section 3.}

The analysis above leads us to consider the possibility that, in fact,
the metrical energy is
not the natural definition of the energy in the AdS/CFT correspondence
context. Thus, we are forced to look for a new definition of the energy, 
and
a brief analysis shows that the natural candidate is the `conserved'
charge which is constructed out of the Noether current corresponding to
time displacements. This is so because the new `conserved' charge,
which from now on we call the `canonical energy', is sensitive to the
addition of boundary terms to the
action, a property inherited from the Noether currents.

The choice of the canonical energy to describe the energy of the bulk
theory may seem strange because, in general, only the metrical energy is
employed. This happens because the stress-energy tensor
Eq.(\ref{9019}) has
some
nice properties, namely, that it is symmetric and it is also traceless for
Weyl-invariant theories (the critical coupling coefficient is
$\varrho =\frac{d-1}{4d}$). However, in principle, we could find physical
systems for which the canonical energy acquires relevance. As an
example of this, we mention black holes physics, in whose context
some aspects of the relation between the metrical
and canonical energies have been analyzed in \cite{fursaev2}. There, it
has been established that both energies are related to different
physical properties of the
system. In the case of the metrical energy, it was shown that it is the
physical energy which enters the first law of black hole
thermodynamics. In the case of the canonical energy, the result in
\cite{fursaev2} is that it is the generator of the time evolution. This is
the reason why the author of \cite{fursaev2} calls the metrical and
canonical
energies as the `energy' and the `Hamiltonian', respectively. Here, we
will
not attempt to perform, for the case of AdS spaces, a formal
demonstration of the statement that the canonical energy is the generator
of the time evolution. We just mention that, on
computing the Poisson brackets, the presence of a boundary requires the
addition of some surface terms to
assure that the Jacobi identity is satisfied \cite{soloviev}\cite{bering}.

One of the purposes of this paper is to present some evidence in
support of our conjecture that the canonical energy is the natural
definition of the energy in the AdS/CFT correspondence context. In order
to do this, we start
by constructing the canonical energy. We know that it is
sensitive to the addition of boundary terms to the action. The origin of
boundary terms in the action is due to the variational principle. In order
to have a stationary action boundary terms, which will depend on the
choice of the boundary conditions, must be introduced. An example of this
is the Gibbons-Hawking term \cite{gibbons} that is added to the
Einstein-Hilbert action in order to make it stationary when the metric
is fixed at the boundary. Other examples
can be found in the AdS/CFT correspondence context, where the boundary
terms that make the action stationary have been shown to be the
generating functionals of the boundary CFT's for the cases of the spinor field
\cite{henneaux2}\cite{frolov1} and the Self-Dual model \cite{our}. But, if
as a first step, we need to construct the canonical energy, and it depends
on the boundary conditions, then we must consider all possible boundary
conditions in order to compute all possible canonical energies of the
system. This leads us to analyze Dirichlet, Neumann and mixed boundary
conditions on the bulk scalar field. As we will show, an interesting
feature about mixed boundary
conditions is that they form a one-parameter family of boundary
conditions. Thus, there is an infinite number of mixed boundary
conditions. In this article, we will develop a new formalism in which we
start from the usual actions for both the minimally and non-minimally
coupled
scalar fields (see Eqs.(\ref{9007}, \ref{9016})). Then, we
will add surface terms to such actions in order to make them stationary
under Dirichlet, Neumann and mixed boundary conditions on the scalar
field. On doing this, we will also keep an eye on the requirement that
the actions must also be stationary under variations of the metric
which vanish at the border. In particular, for mixed boundary
conditions we will find a one-parameter family of surface terms. 

Then, for each resulting action, we will
compute the Noether current corresponding to time displacements in
global coordinates and show
that it is conserved, in the sense that its covariant divergence
vanishes. In the particular case of mixed boundary conditions, there will
be real parameters labelling the corresponding surface terms, and we will 
show that there is a deep analogy between the way in which 
the coupling coefficient $\varrho$ in Eq.(\ref{9016}) is related to
the addition of an `improvement' term to the stress-energy tensor (see
Eq.(\ref{9019})), and the way in which the parameters of mixed boundary
conditions are related to the addition of new `improvement' terms to the
Noether current. Note that, in general, the Noether current will be
different from the usual choice of Eq.(\ref{9020}). It will give rise
to a `conserved' charge which is the canonical energy that we have
discussed before. Then, we will perform a consistent quantization of
the scalar field theory on AdS spaces in global coordinates. This 
quantization is, in
many respects, analogous to the one performed in
\cite{freedman}\cite{freedman8}\cite{mezincescu}. The difference is that, 
in this case, we will carry out the calculations making use of the
canonical energy instead of the metrical one. In particular, we will find 
again regular and irregular modes for
which the canonical energy is conserved, and a key result will be that
for such solutions, and only for them, the canonical energy is also
positive and
finite. This will lead us to obtain a new consistent quantization of the
scalar field theory on AdS. We remark that the
Breitenlohner-Freedman bound Eq.(\ref{9011''}), together with the
constraint Eq.(\ref{12000}) for which irregular modes propagate, 
still hold in this formalism. We will show 
that the important difference between the quantizations which make use of
the metrical and canonical energies is that the constraint 
Eq.(\ref{9041}), for which the metrical energy is conserved,
positive and 
finite for irregular modes propagating in the bulk, is replaced by new
constraints to be found later. A key result that we will also show is 
that, unlike the usual constraint Eq.(\ref{9041}), these new
constraints can be reproduced by means of a generalized version of
the AdS/CFT prescription Eqs.(\ref{9003}, \ref{9013}) which we will
introduce below.

Once we have performed the quantization of the scalar
field theory on
AdS in global coordinates, the next step will be to establish whether the
new results can be mapped through
the AdS/CFT correspondence. An important observation to be made is
that the
Dirichlet, Neumann and mixed boundary conditions which are considered on 
analyzing Noether currents in global coordinates play a fundamental
role in our generalized AdS/CFT prescription, because they fix,
in a natural way, the source
which couples to the boundary conformal operators. It means that, instead
of the usual Dirichlet prescription

\beq
\exp\left( -I _{AdS}[\phi_{0}]\right) \equiv \left<\exp\left(\int d^{d}x
\;
{\cal O}(\vec{x}) \; \phi_{0}(\vec{x})\right)\right>,
\label{9043}
\eeq
the considerations above lead us to employ a generalized AdS/CFT
prescription of the form

\beq
\exp\left( -I _{AdS}[A_{0}]\right) \equiv \left<\exp\left(\int d^{d}x \;
{\cal O}(\vec{x}) \; A_{0}(\vec{x})\right)\right>,
\label{9044}
\eeq
where the source $A_{0}$ which couples to the boundary conformal operator
depends on the boundary conditions. It means that for Dirichlet, Neumann
and mixed boundary conditions the source $A_{0}$ will be the field, its
Lie derivative along the normal vector to the boundary, or a combination
of both of them,
respectively, computed at the border. Thus, the concept of canonical
energy
that we considered on performing the quantization of the scalar field
leads us, in a natural way, to
a generalization of the usual Dirichlet prescription Eq.(\ref{9043}). The
analysis of this generalized
prescription has been carried out in \cite{our2} for the particular
minimally coupled case, and we show in this
paper that the root of such a generalized formalism can be found on
the concept of canonical energy. We stress the fact that, in
considering the non-minimally coupled case, we will
find new results which have not been considered in \cite{our2}. 

A final analysis that we will perform in this article regards a
generalized Legendre transform prescription which does not present the
difficulties mentioned earlier. We also aim to integrate such a
Legendre transform into the generalized prescription Eq.(\ref{9044}) in
such a way that there exists an exact agreement between
the resulting AdS/CFT prescription and the quantization that we develop in
this paper.

Note that the usual Legendre transform prescription
Eq.(\ref{9013}) involves a procedure in which one first expands the
on-shell action in powers of the distance to the boundary, then selects
the leading non-local term, i.e. the generating functional
Eq.(\ref{9012}), and
only then performs the Legendre transformation Eq.(\ref{9013}). This means
that, in the usual formulation along the lines of \cite{witten2}, from all
local
and non-local terms of the on-shell action only the leading non-local
term is taken into account to perform the Legendre transformation. At
first sight, it may seem natural to do so, since the non-leading terms
vanish when the action is taken to the
boundary, and the divergent local terms are of no
relevance on
computing the non-local part of the boundary CFT two-point
function. However, we claim in this work that, even when local terms do
not need
to be taken into account in computing the original generating functional,
they, in general, contain information about the Legendre transformed one,
and thus they
have to be taken into account when performing the Legendre
transformation. In other words, the key observation to be made is that the
operations of expanding the on-shell
action in powers of the distance to the boundary and then select the
leading non-local term, and of performing the Legendre transformation, are
not commuting operations. This means that we have to consider a new
prescription
in which we start by performing the Legendre transformation on the whole
on-shell action, and only then we select the transformed generating
functional, unlike the usual prescription in \cite{witten2} in which
the inverse procedure is considered. We will show that, in general, this
generalized prescription gives rise to results which are different from
those in \cite{witten2}, but that are in complete agreement with the
results arising from the new quantization of the scalar field
theory on the AdS bulk. We will also show that the constraints for which
the irregular modes propagate in the bulk are the same constraints for
which the divergent local terms of the on-shell action vanish. In this
situation, the addition of counterterms is not required.

The generalized prescription for the Legendre transform makes use, in
fact, of the standard form of the Legendre transformation, and
it can be written schematically as

\beq
{\tilde I}_{AdS}[A_{0},{\tilde A}_{0}]\; = I_{AdS}[A_{0}]\; -\;
\int\frac{d^{d}k}{(2\pi)^{d}}\;
A_{0}\left(\vec{k}\right){\tilde A}_{0}\left(-\vec{k}\right)\; .
\label{9045}
\eeq 
Note that, unlike the usual Legendre transform prescription
Eq.(\ref{9013}), where the Legendre transformation acts only on the
leading
non-local term $S[\phi_{0}]$, in the Legendre transformation above
we transform the whole on-shell action
$I_{AdS}[A_{0}]$. We stress the fact that, in general, this
Legendre transformation gives rise to
generating functionals of the boundary CFT's which are different from 
the one that is obtained through the usual Legendre transform prescription
Eq.(\ref{9013}). The Legendre transform of $I_{AdS}[A_{0}]$
is given by the
minimum of ${\tilde I}_{AdS}[A_{0},{\tilde A}_{0}]$ with respect to
$A_{0}$ (for fixed ${\tilde A}_{0}$). An important observation is that the
Legendre transformation Eq.(\ref{9045}) not only modifies the usual
Legendre
transform prescription by acting on the whole on-shell action, but also 
acts on the much more general source $A_{0}$, as required before. The
inclusion
of the generic source $A_{0}$ realizes, in a natural way, the
integration of the Legendre transformation Eq.(\ref{9045}) into the
generalized
AdS/CFT prescription Eq.(\ref{9044}), whose corresponding `conjugated'
prescription reads

\beq
\exp\left( -{\tilde I}_{AdS}[{\tilde A}_{0}]\right) \equiv
\left<\exp\left(\int
d^{d}x \;
{\tilde {\cal O}}(\vec{x}) \; {\tilde A}_{0}(\vec{x})\right)\right>,
\label{9044'}
\eeq
where ${\tilde {\cal O}}$ is the new conformal operator which
couples to the Legendre transformed source ${\tilde A}_{0}$. Note that,
for each
boundary condition, there will be two generating functionals, namely those
arising from the generalized prescriptions
Eqs.(\ref{9044}, \ref{9044'}). In both cases, we will also have
additional local and non-local terms. We remark that the Legendre 
transformation Eq.(\ref{9045}) does not impose any
constraints on $\nu$ or $\varrho$. The role of the Legendre 
transformation is to interpolate between the boundary CFT's
corresponding to regular and irregular modes propagating in the
bulk. On the other hand, we will show that the constraints for which
irregular modes
propagate in the bulk are fixed, in a natural way, by the
generalized AdS/CFT prescription Eqs.(\ref{9044}, \ref{9044'}), and thus
they are not fixed `by hand', as it happens in the usual formalism. In the
cases for which only regular modes
propagate in the bulk, we will show that the Legendre transformation
Eq.(\ref{9045}) interpolates between self-conjugated theories with
conformal
dimension $\Delta_{+}(\varrho)$, as expected. On the other hand, for the
cases for which the irregular modes can propagate in the bulk too, we will
show that the Legendre transformation Eq.(\ref{9045}) interpolates between
the
conformal
dimensions $\Delta_{+}(\varrho)$ and $\Delta_{-}(\varrho)$. In some cases,
the conformal dimension $\Delta_{+}(\varrho)$ belongs to the original
generating functional, and then the conformal dimension
$\Delta_{-}(\varrho)$
arises from the corresponding Legendre transformed
generating functional. However, another new property of this
generalized
formalism
is that, as we will show, there also exist cases for which it is the
conformal dimension $\Delta_{-}(\varrho)$, instead of the usual one 
$\Delta_{+}(\varrho)$, that arises from the
original generating functional, and then the conformal
dimension $\Delta_{+}(\varrho)$ is obtained from the Legendre
transformed generating functional, as expected. It is important to
stress that the cases for which the Legendre transformation 
Eq.(\ref{9045}) interpolates
between different conformal dimensions are precisely the
same for which the divergent local terms of the on-shell action
cancel
out. In this situation, the addition of counterterms is not
required. These results lead us to conclude that divergent terms contain
information
about the Legendre transformed generating functional, and thus they have
to be taken into account.

Note that, unlike the usual Legendre
transform prescription Eq.(\ref{9013}), the Legendre transformation
Eq.(\ref{9045}) can also be successfully employed to compute the
transformed functional in the cases of $\nu$
integer and in particular of $\nu =0$. This follows from
the fact that the expansion in powers of the distance to the
boundary is done only after having carried out the Legendre
transformation.

Another important observation to be made is that the generalized
Legendre transform prescription Eq.(\ref{9045}) has no coefficient to be
fixed `by hand'. In this extended
formalism, the conjugated field of $A_{0}$ is just ${\tilde A}_{0}$,
and its precise form is fixed in a natural way by the generalized
Legendre transform prescription.

In general, we will show that when considering the
generalized AdS/CFT prescription Eqs.(\ref{9044}-\ref{9044'}), we remove 
all the difficulties that arise from the usual prescription
Eqs.(\ref{9013},
\ref{9043}). In particular, we will show that the new constraints for
which irregular
modes propagate in the bulk are fixed in a natural way by the
generalized AdS/CFT prescription Eqs.(\ref{9044}-\ref{9044'}), and thus
they need no longer to be fixed `by hand', as it is the case
when considering the usual formalism. Another point that we will show is
that
such new constraints are the same for which the divergent local terms of
the on-shell action cancel out. In this case, we do not
need to add any counterterms.

Summarizing, in this work we are considering two formulations which, at
first sight, seem to be completely different of each other. One of them is
performed in global
coordinates, and it constructs the canonical energies and requires that 
they are conserved, positive and 
finite. The another one is performed in Poincar\'e coordinates, and it
finds
the boundary CFT two-point functions by making use of the generalized
AdS/CFT prescription. The fact that, as we will show, these two
formulations give rise to exactly 
the same constraints for which irregular modes propagate in the bulk, can
be
considered as
a remarkable non-trivial result, and a strong evidence in support of our
conjectures that the canonical energy is the natural one to be considered
in the AdS/CFT correspondence context, and that the generalized AdS/CFT
prescription Eqs.(\ref{9044}-\ref{9044'}) is needed in order to map to
the border all the information contained in the bulk.

In the following sections, we will present our results in detail. In
particular, we will
construct the canonical energies and perform the quantization of the
scalar field theory on AdS. We will also compute the
boundary CFT's two-point functions by making use of the generalized
AdS/CFT prescription 
Eqs.(\ref{9044}-\ref{9044'}), and show that all the difficulties
that arise from the usual AdS/CFT prescription
Eqs.(\ref{9013}, \ref{9043}) can be removed in this new context. And
finally, in Section 4, we will show that our results
arising from all calculations are in complete agreement, with only one
remarkable exception, namely the case in which the conformal dimension
of the boundary CFT reaches the unitarity bound for the scalar field
theory, and becomes independent of the
effective mass. This is a very puzzling phenomenon which has no bulk
counterpart.

\section{The Canonical Energy}

In this section, we formulate Dirichlet, Neumann and mixed
boundary-value problems for a real massive scalar field theory
on $AdS_{d+1}$ in both minimally and non-minimally coupled cases. Then,
for each one of these boundary-value problems, we compute the
Noether current corresponding to time displacements in global
coordinates, and show that it is conserved, in the sense that
its covariant divergence vanishes. By requiring that the corresponding
`conserved' charge, i.e. the canonical energy, is actually conserved, we
will find regular and irregular 
solutions, and new constraints for which the irregular ones propagate in
the bulk. We will
also show that for such regular and irregular modes,
and only for them, the canonical energy is also positive and finite,   
thus leading to a consistent quantization of the scalar field theory on
AdS. We leave for the next section to show that, contrary to what
happens to the constraint Eq.(\ref{9041}), which arises on
considering the metrical energy, the new constraints for which irregular
modes propagate in the bulk can be reproduced through a generalized
AdS/CFT prescription.

We consider a d+1 dimensional space ${\cal M}$ with metric
$g_{\mu\nu}$ of `mostly plus' signature. Let $x^{\mu}$ be coordinates on
${\cal M}$. We begin by analyzing the non-minimally coupled case, and
leave the minimally coupled case to be considered later. The usual
matter contribution to the action of a system containing a
massive real scalar field non-minimally coupled to the metric $g_{\mu\nu}$
is given by

\beq
I_{0}=-\frac{1}{2}\;\int_{\cal M} d^{d+1} x \;\sqrt{g}\;
\left[
g^{\mu\nu}\partial_{\mu}\phi\;\partial_{\nu}\phi\; +
\;\left( m^{2}+\;\varrho R\right)\phi^{2}\right] \; ,
\label{301}
\eeq
where $m$ is the mass of the scalar field, $R$ is the Ricci scalar
corresponding to $g_{\mu\nu}$ and $\varrho$ is an arbitrary coupling
coefficient between the scalar field and the metric.

We consider ${\cal M}$ as foliated by a one-parameter family of d 
dimensional timelike surfaces $\partial{\cal M}_{\rho}$ homeomorphic to
the
timelike boundary $\partial{\cal M}$. Here $\rho$ is a real parameter and
in
particular we assume that $\partial{\cal M}_{\rho}\rightarrow
\partial{\cal M}$ as $\rho\rightarrow\rho_{0}$. We refer
to $\partial{\cal M}_{\rho}$ as the boundary to the interior region
${\cal M}_{\rho}$. In general, the limit $\rho\rightarrow\rho_{0}$ will be
taken only at the end of the calculations.

It can be shown that, in order for the action $I_{0}$ to be 
stationary under an infinitesimal variation of the metric

\beq
g_{\mu\nu}\rightarrow g_{\mu\nu} + \delta g_{\mu\nu}\; ,
\label{701}
\eeq
the metric and certain of its normal derivatives must fixed at
$\partial{\cal M}_{\rho}$. In order to have a well-defined variational
principle, we must add to $I_{0}$ a surface term of the form
\cite{barvinsky}

\beq
\varrho\int_{\partial{\cal M}_{\rho}} d^{d}x\;\sqrt{h}\; K\;\phi^{2}\; ,
\label{702}
\eeq
which accounts for the terms of the variation containing derivatives of
the metric. Here 
$h_{\mu\nu}$ is the induced metric on $\partial{\cal M}_{\rho}$, and 
$K$ is the trace of the extrinsic curvature. It is given by 

\beq
K=-\nabla_{\mu}n^{\mu}\; ,
\label{705}
\eeq
where $n_{\mu}$ is the outward pointing unit normal
vector to $\partial{\cal M}_{\rho}$.
The surface term Eq.(\ref{702}) is just the natural extension of the
Gibbons-Hawking term \cite{gibbons} which is added to the Einstein-Hilbert
action in order to have a well-defined variational principle. 

Then, the resulting matter contribution to the action reads

\beq
I_{D}=-\frac{1}{2}\;\int_{{\cal M}_{\rho}} d^{d+1} x \;\sqrt{g}\;
\left[ g^{\mu\nu}\partial_{\mu}\phi\;\partial_{\nu}\phi\; +
\;\left( m^{2}+\;\varrho R\right)\phi^{2}\right]
\; +\;\varrho\int_{\partial{\cal M}_{\rho}}
d^{d}x\;\sqrt{h}\; K\;\phi^{2}\; .
\label{714}
\eeq
The reason for the notation `$I_{D}$' to refer to the matter
contribution will be clarified later. It can be shown that,
as expected, $I_{D}$ is stationary under the variation
Eq.(\ref{701}) when the metric is fixed at the boundary.

So far, we have concentrated only on infinitesimal variations of the
metric, as in Eq.(\ref{701}). In order to consider boundary-value problems
on the scalar field, we have to analyze the
transformation properties of the action under infinitesimal variations of
the scalar field as follows

\beq
\phi\rightarrow\phi +\delta\phi\; .
\label{716}
\eeq
Under the variation above, the matter contribution $I_{D}$ 
transforms as

\beq
\delta_{\phi} I_{D}=-\int_{\partial{\cal M}_{\rho}}
d^{d}x\;\sqrt{h}\;\left(
\partial_{n}\phi - 2\varrho\; K \phi\right)\;\delta
\phi\; ,
\label{717}
\eeq  
where $\partial_{n}\phi$ is the Lie derivative of $\phi$ along $n_{\mu}$, 
and it is given by

\beq
\partial_{n}\phi = n^{\mu}\partial_{\mu}\phi\; .
\label{718}
\eeq
Note that, in Eq.(\ref{717}), the absence of a bulk
contribution is due to the equation of motion

\beq
\nabla^{2}\phi - \left(m^{2}+\varrho R\right)\phi =0\; .
\label{718'}
\eeq

The variation Eq.(\ref{717}) shows that the action
$I_{D}$ is stationary for a Dirichlet boundary condition which
fixes
the value of the scalar field $\phi$ at $\partial{\cal M}_{\rho}$, namely

\beq
\delta\phi\; {\huge\mid}_{\partial{\cal
M}_{\rho}}=0\qquad\qquad\mbox{(Dirichlet)}\; .
\label{719}
\eeq
This is the reason why we called this action `$I_{D}$'.\footnote{Note
that we could also have considered the boundary condition $\left(
\partial_{n}\phi - 2\varrho K
\phi\right){\huge\mid}_{\partial{\cal
M}_{\rho}}=0$. However, this kind of boundary condition plays no role in
the
present formalism. Analogous observations hold for the remaining
boundary conditions to be considered in this paper.}

In this way, we have found an action which has a well-defined variational
principle when both the metric and the scalar field are fixed at the
boundary. However, for more general boundary conditions, new surface
terms must be added to the matter contribution $I_{D}$. On doing
this, we define the following action

\beq
I_{M,I}=I_{D} + \int_{\partial{\cal M}_{\rho}}
d^{d}x\;\sqrt{h}\;\phi\;\partial_{n}\phi\; ,
\label{720}
\eeq
which differs from $I_{D}$ only by a surface term. It can be
shown that the new surface term does not spoil the property of
having a well-defined variational principle under Eq.(\ref{701}) when the
metric is fixed at
the boundary. Besides, it can be verified that, under the variation of
the scalar field Eq.(\ref{716}), the action $I_{M,I}$ transforms
as

\beq
\delta_{\phi}I_{M,I} = \int_{\partial{\cal M}_{\rho}}
d^{d}x\;\sqrt{h}\;\phi\;\delta\psi^{I}\; ,
\label{722}
\eeq
where the absence of a bulk contribution is due to the equation of
motion. The field $\psi^{I}$ in the equation above is
defined as

\beq
\psi^{I} = \partial_{n}\phi + 2\varrho K\phi\; .
\label{723}
\eeq
Then, the action $I_{M,I}$ is stationary under a boundary
condition which fixes $\psi^{I}$ at $\partial{\cal M}_{\rho}$, and we
call this as `Type I' mixed boundary condition. Namely

\beq
\delta\psi^{I}{\huge\mid}_{\partial{\cal
M}_{\rho}}=0\qquad\qquad\mbox{(Type
I mixed)}\; .
\label{724}
\eeq
Note that there exists a one-parameter family of Type I mixed boundary
conditions.

So far, we have considered the non-minimally coupled case. For the
minimally coupled case $\varrho =0$, the Dirichlet boundary condition
Eq.(\ref{719}) stays as a Dirichlet boundary condition. On the other hand,
for $\varrho =0$ the Type I mixed boundary condition in 
Eq.(\ref{724}) becomes a Neumann boundary condition, as noted from
Eq.(\ref{723}). Thus, at first sight, it seems that for the minimally
coupled case only Dirichlet and Neumann boundary conditions are
possible. However, this is not the case. For the particular case 
of $\varrho =0$, consider the following action

\beq
I_{M,I\!I}=I_{D}{\huge\mid}_{\varrho =0}-
\;\lambda\int_{\partial{\cal M}_{\rho}}
d^{d}x\;\sqrt{h}\;\left(\partial_{n}\phi\right)^{2}\; ,
\label{725}
\eeq
where $\lambda$ is an arbitrary real coefficient. The action
$I_{M,I\!I}$ differs from
$I_{D}{\huge\mid}_{\varrho =0}$ only on a
surface term, and to the best of our knowledge it has been considered for
the first time in \cite{our2},
in the context of the AdS/CFT correspondence. As in the cases
considered before, it can be shown that the new surface term does not
spoil the property of having a well-defined variational principle
under Eq.(\ref{701}) when the
metric is fixed at the boundary. Furthermore, under the variation of
the scalar field Eq.(\ref{716}) the action $I_{M,I\!I}$ transforms
as

\beq
\delta_{\phi} I_{M,I\!I} = -\int_{\partial{\cal M}_{\rho}}
d^{d}x\;\sqrt{h}\;\partial_{n}\phi\;\delta\psi^{I\!I}\; ,
\label{727}
\eeq
where, as before, the absence of a bulk contribution is due to the
equation of motion. The field $\psi^{I\!I}$ in
Eq.(\ref{727}) is defined as

\beq
\psi^{I\!I} = \phi + 2\lambda\;\partial_{n}\phi\; .
\label{728}
\eeq
Thus, the action $I_{M,I\!I}$ is stationary under a boundary
condition which fixes $\psi^{I\!I}$ at $\partial{\cal M}_{\rho}$, and we
call
this as `Type II' mixed boundary condition. Namely
  
\beq
\delta\psi^{I\!I}{\huge\mid}_{\partial{\cal
M}_{\rho}}=0\qquad\qquad\mbox{(Type II mixed)}\; .
\label{729}
\eeq
Note that, as in the case of Type I mixed boundary conditions, there
exists a one-parameter family of Type II mixed boundary conditions. In the
case of Type I mixed boundary conditions, the corresponding parameter is
just the coupling coefficient between the field and the metric $\varrho$,
whereas for Type II mixed boundary conditions the
corresponding parameter is $\lambda$, which plays for the Type II mixed
boundary conditions a role which is analogous to the one played by the
coupling coefficient $\varrho$ for Type I mixed boundary conditions.

In this way, we have computed the actions for scalar field theory on a
curved
space which correspond to Dirichlet, Neumann and mixed boundary conditions 
on $\partial{\cal M}_{\rho}$ in both the minimally and non-minimally
coupled cases, and that have also a
well-defined variational principle under Eq.(\ref{701}) when the metric is
fixed at the
boundary. 

From now on, we concentrate on the specific case of d+1 dimensional 
Anti-de Sitter spaces, whose metric in global coordinates reads

\beq
ds^{2} =
\frac{1}{cos^{2}\rho}\;(-d\tau^{2}+d\rho^{2}+sin^{2}\rho\;
d\Omega_{d}^{2})\quad (d\geq 2)\; ,
\label{39}
\eeq
where we have fixed the radius of $AdS_{d+1}$ equal to one. Here,
$d\Omega_{d}^{2}$ is the angular element, and $\rho$ and $\tau$ are the
radial and time coordinates respectively. They satisfy

\ba
&& \quad 0\leq\rho<\frac{\pi}{2}\qquad (d\geq 2)\; ,
\label{36}
\\
&& -\pi\leq\tau<\pi\; .
\label{35}
\ea
In order to avoid closed timelike curves, we follow \cite{avis} and pass
to the universal covering space CAdS by setting

\beq
-\infty<\tau<\infty\; .
\label{38}
\eeq

We consider that the boundary-value problems analyzed before are
formulated on surfaces
of fixed radial coordinate $\rho$. Such surfaces are homeomorphic to the
boundary at $\rho\rightarrow\rho_{0}=\frac{\pi}{2}.$ The surface forming
an outer normal vector is given by

\beq
n_{\mu}\; =\;\frac{1}{cos\rho}\;\delta^{(\rho)}_{\mu}\; ,
\label{40'}
\eeq
and then we find

\beq
K=-\frac{1}{sin\;\rho}\;\left(d-cos^{2}\rho\right)\; .
\label{100}
\eeq

The first step to perform the quantization of the scalar 
field theory on AdS is to compute
the Noether currents corresponding to time displacements. In order to do
this, we consider a Killing vector $\xi^{\mu}$ of the form

\beq
\xi^{\mu}=-\delta^{\mu}_{(\tau)}\; ,
\label{40}
\eeq
where we have included a minus sign because of the ``mostly plus''
signature of the metric. Then, we perform the isometry

\beq
x^{\mu}\longrightarrow x^{\mu} + \sigma\;\xi^{\mu}\; ,
\label{731}
\eeq
where $\sigma$ is an infinitesimal constant parameter. The variations of
the actions $I_{D}$, $I_{M,I}$ and $I_{M,I\!I}$
under the isometry above are of the generic form

\beq
\delta_{\xi} I=\sigma\int_{\partial{\cal M}_{\rho}}
d^{d}x\;\sqrt{h}\;n_{\mu}\; J^{\mu}\; ,
\label{733}
\eeq
where $J^{\mu}$ is the Noether current, and since Noether currents are
sensitive to
the addition of boundary terms to the action, we expect to have different
Noether currents corresponding to the actions $I_{D}$, $I_{M,I}$ and
$I_{M,I\!I}$. Making use of the equation of motion
Eq.(\ref{718'}), and simplifying the calculations by noting that
$g^{\mu\nu}$, $R$, $n_{\mu}$ and $K$ are invariant under 
Eq.(\ref{731}),\footnote{Note that $n^{\mu}$ is not a true
vector, and its variation includes an extra deviation term.} we find the
following
Noether currents corresponding to the time displacement Eq.(\ref{731})

\ba
J^{\mu}_{D} &=& -\Theta^{\mu}_{\;\;\tau}-
\varrho\left[
\delta^{\mu}_{\;\tau}\nabla_{\nu}\left(K n^{\nu} 
\phi^{2}\right)-K n^{\mu}\partial_{\tau}\phi^{2}\right]
\; ,
\label{736}
\\ \nonumber
\\
J^{\mu}_{M,I} &=& -\Theta^{\mu}_{\;\;\tau}-
\varrho\left[
\delta^{\mu}_{\;\tau}\nabla_{\nu}\left(K n^{\nu}
\phi^{2}\right)-K n^{\mu}\partial_{\tau}\phi^{2}\right]
\nonumber\\ && \qquad\;\; -\;\frac{1}{2}
\left[\delta^{\mu}_{\;\tau}\nabla^{2}
\phi^{2}-\;\partial^{\mu}
\partial_{\tau}\phi^{2}\right]\; ,
\label{737}
\\ \nonumber
\\
J^{\mu}_{M,I\!I} &=&
-\Theta^{\mu}_{\;\;\tau}{\huge\mid}_{\varrho =0}-\lambda \;[\;
n^{\nu}\partial^{\mu}\phi \;\partial_{\nu}
\partial_{\tau}\phi+
n^{\nu}\partial_{\nu}\phi\;\partial^{\mu}
\partial_{\tau}\phi\nonumber\\
&&\qquad\qquad\qquad\quad -\delta^{\mu}_{\;\tau}\nabla_{\nu}\left(
n^{\alpha}\partial_{\alpha}\phi\;\partial^{\nu}\phi\right)\;]\; , 
\label{738}
\\ \nonumber
\\
\Theta_{\mu\nu} &=& \partial_{\mu}\phi\;\partial_{\nu}\phi\;-\;\frac{1}{2}
\; g_{\mu\nu}\left[
g^{\alpha\beta}\partial_{\alpha}\phi\;\partial_{\beta}\phi\;
+\left(m^{2}+\varrho R\right)\phi^{2}\right]\; .
\label{739}
\ea
Making use of the equation of motion, we can show that the above Noether
currents
are conserved, as expected. Note that the surface terms in the actions
$I_{D}$, $I_{M,I}$ and $I_{M,I\!I}$ introduce `improvement' terms to the
Noether currents which
are analogous to the `improvement' term that the coupling between the
scalar field and the metric introduces to the stress-energy tensor (see
Eq.(\ref{9019})). The canonical energies are obtained from the $J^{\tau}$
component of the Noether currents Eqs.(\ref{736}-\ref{738}), and they read

\ba
E_{D} &=& -\int d^{d}x\;\sqrt{g}\;\left[
\Theta^{\tau}_{\;\; \tau} +\;\varrho\;\nabla_{\mu}
\left(K n^{\mu}\phi^{2}\right)\right]\; , 
\label{41}\\
E_{M,I} &=& -\int d^{d}x\;\sqrt{g}\;\left[
\Theta^{\tau}_{\;\; \tau} +\;\varrho\;\nabla_{\mu}
\left( K n^{\mu}\phi^{2}\right)
+\;\frac{1}{2}\;\left(\nabla^{2}\phi^{2}- 
\partial^{\tau}\partial_{\tau}\phi^{2}\right)\right]\; ,
\label{41'}\\
E_{M,I\!I} &=& -\int d^{d}x\;\sqrt{g}\;\{\
\Theta^{\tau}_{\;\; \tau}\;{\huge\mid}_{\varrho =0}\;
+\;\lambda\; \left[\;\partial_{\tau}
\left(\partial^{\tau}\phi\; \partial_{n}\phi\right)
-\;\nabla_{\mu}
\left(\partial^{\mu}\phi\;\partial_{n}\phi\right)\;\right
]\;\}\
\; ,\nonumber\\
\label{41''}
\ea
where the integration is carried out over the spatial coordinates. 

The next step is to solve the equation of motion Eq.(\ref{718'}) in global
coordinates. This has been done in 
\cite{kraus}\cite{freedman}\cite{freedman8}\cite{mezincescu}\cite{avis}, 
and we
recall here the main results in order to introduce some notation and
make some relevant comments. The solution can be expanded in modes as

\beq
\phi=\sum_{\omega l\{ m\} }\left[ a_{\omega l\{ m\}}\phi_{\omega l\{ m\}}
+a_{\omega l\{ m\}}^{*}\phi_{\omega l\{ m\}}^{*}\right]\qquad (d\geq
3)\; ,
\label{42}
\eeq
where the $a_{\omega l\{ m\}}$'s are complex coefficients and the
modes $\phi_{\omega l\{ m\}}$ are of the form

\beq
\phi_{\omega l\{ m\}}=N_{\omega
l}\;e^{-i\omega\tau}Y_{l\{m\}}(\Omega_{d})\;
G_{\omega l}(\rho)\qquad (d\geq 3)\; .
\label{43}
\eeq
Here the $\omega$'s are real numbers to be determined later,
$N_{\omega
l}$ are normalization coefficients and the
$Y_{l\{m\}}$'s are spherical harmonics in ${\bf S}^{d-1}$. We will analyze
here the case $d\geq 3$, but
the case $d=2$ is analogous and we will comment on it 
later in this section. 

The radial part of the equation of motion reads

\beq
\partial_{\rho}\left[(tan\;\rho)^{d-1}\partial_{\rho}
G_{\omega l}(\rho)
\right]=\left[\frac{M^{2}(\varrho)}{cos^{2}\rho}
-\omega^{2}+l(l+d-2)\frac{1}{sin^{2}\rho}\right](tan\;\rho)^{d-1}
G_{\omega l}(\rho) \; ,
\label{43''}
\eeq
where $M^{2}(\varrho)$ is the effective mass on Eq.(\ref{9016'}). It is
convenient to write

\beq
G_{\omega l}(\rho)=(sin\;\rho)^{l}(cos\;\rho)^{\Delta}F_{\omega
l}(\rho)\; ,
\label{43'}
\eeq
where $\Delta$ is an arbitrary real parameter. We choose

\beq
\Delta (\Delta - d)=M^{2}(\varrho)\; ,
\label{44}
\eeq
and the solutions of this equation are the coefficients
$\Delta_{\pm}(\varrho)$ in Eq.(\ref{9006''}). After straightforward
algebra, Eq.(\ref{43''}) can be written as

\beq
0=x(1-x)\;\frac{d^{2}F^{\pm}_{\omega
l}}{dx^2}+\left[c-\left(a_{\pm}+b_{\pm}+1\right)\; x\right]
\;\frac{dF^{\pm}_{\omega l}}{dx}-a_{\pm}\; b_{\pm}F^{\pm}_{\omega l}\; ,
\label{47}
\eeq
where

\beq
x=sin^{2}\rho\; ,
\label{48}
\eeq
\ba
a_{\pm}&=&\frac{1}{2}\;\left[ 
l+\Delta_{\pm}(\varrho)-\omega\right]\; ,\nonumber\\
b_{\pm}&=&\frac{1}{2}\;\left[ 
l+\Delta_{\pm}(\varrho)+\omega\right]\; ,\nonumber\\
c &=& l+\frac{d}{2}\; .
\label{49}   
\ea
As pointed out in \cite{freedman}\cite{freedman8}\cite{avis}, for each
choice $\Delta_{+}(\varrho)$
or $\Delta_{-}(\varrho)$ only one of the two independent
solutions of Eq.(\ref{47}) is regular at the origin in $\rho=0$, and it is
given by

\beq
F^{\pm}_{\omega l}(\rho)={_{2}F_{1}}(a_{\pm},b_{\pm};c\; ;sin^{2}\rho)\; ,
\label{50}
\eeq
where ${_{2}F_{1}}(a,b;c\; ;x)$ is the hypergeometric function. The
solution of Eq.(\ref{43''}) which is regular at the
origin\footnote{Note that, at first sight, we are in presence of two
regular solutions of Eq.(\ref{43''}), namely
$(sin\;\rho)^{l}(cos\;\rho)^{\Delta_{+}(\varrho)}F^{+}_{\omega 
l}(\rho)$ and $(sin\;\rho)^{l}(cos\;\rho)^{\Delta_{-}(\varrho)}
F^{-}_{\omega l}(\rho)$. However, Eq.(\ref{43''}) is a second order
differential
equation, and taking into account that we have discarded its non-regular
solution, we expect that only one solution is left. In fact, the
identity $(cos\;\rho)^{\Delta_{+}(\varrho)}F^{+}_{\omega
l}(\rho)=(cos\;\rho)^{\Delta_{-}(\varrho)}F^{-}_{\omega
l}(\rho)$ can be shown by using standard properties of
the hypergeometric function. As far as we know, this was not pointed out 
before.} can be
expanded in powers of $cos\rho$ by making use of the identities presented
in Appendix A. It is important to perform the analysis by considering the
cases of $\nu$ not integer, $\nu$ integer but not zero
and $\nu =0$ separately. After some calculations, we
arrive at the following series expansions of the regular radial solution
$G_{\omega l}(\rho)$.
\\
\\
i) $\;$ For $\nu\not\in {\bf Z}$,
\ba
G_{\omega l}(\rho)&=&(sin\;\rho)^{l}\;(cos\;\rho)^{\Delta_{-}(\varrho)}
\frac{\Gamma(l+\frac{d}{2})\Gamma(\nu)\Gamma(1-\nu)}
{\Gamma(a_{+})\Gamma(b_{+})
\Gamma(a_{-})\Gamma(b_{-})}\nonumber\\  
&\times&
\sum_{n\geq 0}\frac{1}{n!}\;(cos\;\rho)^{2n}
\; \left[\;\frac{\Gamma(n+a_{-})\Gamma(n+b_{-})}   
{\Gamma(n+1-\nu)}-
\frac{\Gamma (n+a_{+})\Gamma(n+b_{+})}
{\Gamma(n+1+\nu)}\;
(cos\;\rho)^{2\nu}\; \right]\; .\nonumber\\
\label{52}
\ea
ii) $\;$ For $\nu =0$,
\ba
G_{\omega l}(\rho)&=&-(sin\;\rho)^{l}\;(cos\;\rho)^{\Delta_{+}(\varrho)}
\frac{\Gamma(l+\frac{d}{2})}   
{\Gamma(a_{+})\Gamma(b_{+})
\Gamma(a_{-})\Gamma(b_{-})}\nonumber\\ &\times&
\sum_{n\geq 0}\frac{1}{(n!)^{2}}\;\Gamma(n+a_{+})\Gamma(n+b_{+})
\left[2\; ln\; cos\;\rho +
u_{n}(a_{-},b_{-},0) \right](cos\;\rho)^{2n}\; ,
\nonumber\\
\label{53}
\ea
where $u_{n}(a,b,m)$ is given by Eq.(\ref{924'}).
\\
\\
iii) $\;$ For $\nu\in {\bf Z}, \;\nu >0$,
\ba
G_{\omega l}(\rho)&=&(sin\;\rho)^{l}\;(cos\;\rho)^{\Delta_{-}(\varrho)}
\frac{\Gamma(l+\frac{d}{2})}
{\Gamma(a_{+})\Gamma(b_{+})
\Gamma(a_{-})\Gamma(b_{-})}\nonumber\\ &\times & {\bigg [}
\sum^{\nu -1}_{n=0}\frac{(-1)^{n}}{n!}\;
\Gamma(\nu -n)\Gamma(n+a_{-})\Gamma(n+b_{-})(cos\;\rho)^{2n}
\nonumber\\ &-&
(-1)^{\nu}(cos\;\rho)^{2\nu}
\sum_{n\geq 0}\frac{\Gamma(n+a_{+})\Gamma(n+b_{+})}{\Gamma (n+1)\Gamma
(n+1+\nu)}\nonumber\\ && \qquad\qquad\qquad\qquad \times
\left[2\; ln\; cos\;\rho +
u_{n}(a_{-},b_{-},\nu) \right](cos\;\rho)^{2n}{\bigg ]}
\; .\nonumber\\
\label{54}
\ea

The next step is to analyze the conditions for 
which the canonical energies
Eqs.(\ref{41}-\ref{41''}) are conserved, finite and positive. As in the
case of the
metrical energy, the canonical energy is actually conserved
if there is no flux of the corresponding conserved current through the
boundary at $\rho\rightarrow\rho_{0}=\frac{\pi}{2}.$ This condition gives
rise to
regular and irregular modes, just as it happens to the usual formalism
involving the metrical energy along the lines of 
\cite{freedman}\cite{freedman8}\cite{mezincescu}. Thus, regular and
irregular modes arise on considering either the metrical and the
canonical energies. The conditions (for $n=0,1,2,\cdots $)

\ba
a_{+} &=& -n,\quad \mbox{or}\quad b_{+}=-n\qquad\mbox{(regular
modes),}\nonumber\\
a_{-}&=& -n,\quad \mbox{or}\quad b_{-}=-n\qquad\mbox{(irregular 
modes),}
\label{170}
\ea
quantize $\omega$ as follows

\ba
\mid\omega^{+}_{n}\mid &=& l+\Delta_{+}(\varrho)+2n\qquad\mbox{(regular
modes),}\nonumber\\
\mid\omega^{-}_{n}\mid &=& l+\Delta_{-}(\varrho)+2n\qquad\mbox{(irregular
modes),}
\label{171}
\ea
and give rise to regular and irregular solutions, whose asymptotic
behavior is of the form

\ba
G^{+}_{nl} &\sim &
{\hat\epsilon}^{\Delta_{+}(\varrho)}\qquad\mbox{(regular
modes),}\nonumber\\
G^{-}_{nl} &\sim&
{\hat\epsilon}^{\Delta_{-}(\varrho)}\qquad\mbox{(irregular
modes),}
\label{172}
\ea
where ${\hat\epsilon} =cos\;\rho$.

The main result of this section will be the derivation of the
constraints for which
the canonical energies Eqs.(\ref{41}-\ref{41''}) are actually conserved,
allowing regular and irregular modes to propagate in the bulk. The
conditions for which the flux of the Noether currents corresponding to
the canonical energies vanish can be written for a superposition of
two modes as:
\\
\\
i) $\;$ For Dirichlet boundary conditions,

\beq
0=lim_{{\hat\epsilon}\rightarrow 0} \left[ \hat\epsilon^{-(d-1)}
\;G_{\omega l}\left(\partial_{\hat\epsilon}G_{\omega '
l}-2\varrho\; d\;{\hat\epsilon}^{-1}G_{\omega 'l}\right)\right ]\; .
\label{55}
\eeq
\\
ii) $\;$ For Type I mixed boundary conditions,

\beq
0=lim_{{\hat\epsilon}\rightarrow 0} \left[ \hat\epsilon^{-(d-1)}
\;G_{\omega l}\left(\partial_{\hat\epsilon}G_{\omega '
l}+2\varrho\; d\;{\hat\epsilon}^{-1}G_{\omega 'l}\right)\right ]\; .
\label{55''}
\eeq
\\
iii) $\;$ For Type II mixed boundary conditions,

\beq
0=lim_{{\hat\epsilon}\rightarrow 0} \left[ \hat\epsilon^{-(d-1)}
\;\partial_{\hat\epsilon}G_{\omega l}\left( G_{\omega'
l}-2\lambda\;{\hat\epsilon}\;\partial_{\hat\epsilon}G_{\omega'
l}\right){\huge\mid}_{\varrho =0}\right ]\; .
\label{56'}
\eeq

As in the cases of the scalar product Eq.(\ref{9005}) and the metrical 
energy which is constructed out of the conserved current Eq.(\ref{9020}),
for the canonical energies the regular modes propagate for any real
$\nu\geq 0$, whereas the irregular ones propagate when the
constraint Eq.(\ref{12000}) is satisfied. There is nothing new
here. Note also that the
Breitenlohner-Freedman bound Eq.(\ref{9011''}) still holds, as
expected. The key result is that, instead of the usual constraint
Eq.(\ref{9041}) corresponding to irregular modes propagating in the bulk
when the metrical energy is considered, for the case of the canonical
energies there arise new constraints which can be computed from the
conditions Eqs.(\ref{55}-\ref{56'}). Making use of
Eqs.(\ref{52}-\ref{54}), we find that the new constraints read as follows:
\\
\\
i) $\;$ Dirichlet boundary conditions.
\\

In this case we find the constraint

\beq
\varrho =\frac{1}{2d}\;\Delta_{-}(\varrho)\; ,
\label{150}
\eeq
which has two possible solutions

\beq
\varrho_{D}^{\pm}=\frac{d-1}{8d}\;
\left[1\pm\sqrt{1+\left(\frac{4m}{d-1}\right)^{2}}\right]\; .
\label{151}
\eeq
\\
ii) $\;$ Type I mixed boundary conditions.
\\

The constraint corresponding to this case reads

\beq
\varrho =-\frac{1}{2d}\;\Delta_{-}(\varrho)\; ,
\label{152}
\eeq
and has solutions

\beq
\varrho_{M,I}^{\pm}=-\frac{3d+1}{8d}\;
\left[1\mp\sqrt{1+\left(\frac{4m}{3d+1}\right)^{2}}\right]\; .
\label{153}
\eeq
\\
iii) $\;$ Type II mixed boundary conditions.
\\

In this case, the constraint reads

\beq
\lambda = \lambda_{M,I\! I}\; ,
\label{154}
\eeq
where

\beq
\lambda_{M,I\! I}=\frac{1}{2\Delta_{-}(0)}\; .
\label{155}
\eeq
The key result that we will show in the next section is that, unlike the
usual constraint Eq.(\ref{9041}) for which the metrical energy is
conserved for irregular modes propagating in the bulk,
the new constraints Eqs.(\ref{150}, \ref{152}, \ref{154}) can be
reproduced in a natural way by means of a generalized
AdS/CFT prescription, which also
contains the information about the constraint Eq.(\ref{12000}). We will
also show in the next section that the constraints Eqs.(\ref{12000}, 
\ref{150}, \ref{152}, \ref{154}) are the ones for which the divergent
local terms of the on-shell actions cancel out. In this situation, the
addition of counterterms is not required.

As a consequence of Eq.(\ref{170}), the regular and irregular modes can
be written in terms of Jacobi polynomials as follows

\beq
G^{\pm}_{nl}(\rho)=(sin\;\rho)^{l}(cos\;\rho)^{\Delta_{\pm}(\varrho)}
\; n!\;\frac{\Gamma\left(l+\frac{d}{2}\right)}
{\Gamma\left(l+\frac{d}{2}+n\right)}\;
P^{(l+\frac{d}{2}-1,\pm\nu)}_{n}(cos\; 2\rho)\; .
\label{64}
\eeq
For regular and irregular modes, the general solution
Eq.(\ref{42}) reads

\beq
\phi^{\pm}=\sum_{n l\{ m\} }\left[ a^{\pm}_{n l\{
m\}}\phi^{\pm}_{n l\{ m\}}
+a^{\pm *}_{n l\{ m\}}\phi^{\pm *}_{n l\{ m\}}\right]\qquad (d\geq
3)\; ,
\label{70}
\eeq
where

\beq
\phi^{\pm}_{n l\{ m\}}=N^{\pm}_{n 
l}\;e^{-i\omega^{\pm}_{n}\tau}Y_{l\{m\}}(\Omega_{d})\;
G^{\pm}_{n l}(\rho)\qquad (d\geq 3)\; .
\label{71}
\eeq
The normalization coefficients $N^{\pm}_{nl}$ can be computed from the
orthogonality properties of the Jacobi polynomials and the spherical
harmonics, and their precise values are not relevant for our present
purposes.

Finally, we will show that for regular and irregular modes and for the
constraints Eqs.(\ref{12000}, \ref{150}, \ref{152}, \ref{154}), the
canonical energies Eqs.(\ref{41}-\ref{41''}) are positive and finite, thus
leading to a consistent quantization of the scalar field theory on
AdS. We define the field $\phi'^{\pm}$ through

\beq
\phi^{\pm}=(cos\;\rho)^{\Delta_{\pm}(\varrho)}\;\phi'^{\pm}\; .
\label{65}
\eeq
This means that $\phi'^{+}$ ($\phi'^{-}$) approaches the boundary as a
constant for regular (irregular) solutions. 

Using Eqs.(\ref{41}-\ref{41''}, \ref{44}, \ref{65}) and integrating by
parts, we find that

\ba
E^{\pm}_{D} &=& E^{\pm}_{V}\; +\; \int
d\Omega_{d}\;(cos\;\rho)^{2
\left(\Delta_{\pm}(\varrho )-1\right)}\; (tan\;\rho)^{d-1}\nonumber\\ 
&&\qquad\qquad\qquad\times\;cos\;\rho\;sin\;\rho 
\left(-\frac{\Delta_{\pm}(\varrho )}{2}\; +
\;\frac{\varrho\; d}{sin^{2}\rho}
-\;\frac{\varrho}{tan^{2}\rho}\right)\left(\phi'^{\pm}\right)^{2}
\; {\bigg |}_{\rho =\frac{\pi}{2}}\; ,\nonumber\\
\label{68'}
\ea
\ba
E^{\pm}_{M,I}&=& E^{\pm}_{V}
\; +\; \int d\Omega_{d}\;(cos\;\rho)^{2 
\left(\Delta_{\pm}(\varrho )-1\right)}\; (tan\;\rho)^{d-1}\nonumber\\
&&\qquad\qquad\qquad\times\; {\bigg [}
cos\;\rho\; sin\;\rho\;
\left(\frac{\Delta_{\pm}(\varrho)}{2}\; +
\;\frac{\varrho\; d}{sin^{2}\rho}
-\;\frac{\varrho}{tan^{2}\rho}\right)
\left(\phi'^{\pm}\right)^{2}\nonumber\\
&&\qquad\qquad\qquad\qquad -\;cos^{2}\rho\;\phi'^{\pm}
\partial_{\rho}\phi'^{\pm} {\bigg ]}
\; {\bigg |}_{\rho =\frac{\pi}{2}}
\; ,\quad\qquad\qquad\qquad\qquad\qquad\qquad
\label{68''}
\ea
\ba
E^{\pm}_{M,I\! I} &=& E^{\pm}_{V}{\huge\mid}_{\varrho =0}\;+\; \int  
d\Omega_{d}\;(cos\;\rho)^{2
\left(\Delta_{\pm}(0)-1\right)}\; (tan\;\rho)^{d-1}\nonumber\\
&&\qquad\qquad\qquad\times\;{\bigg [}\lambda\;cos^{3}\rho\;
\left(\partial_{\rho}\phi'^{\pm}\right)^{2}-2\;\lambda\;
\Delta_{\pm}(0)\; cos^{2}\rho\; sin\;\rho
\;\phi'^{\pm}\partial_{\rho}\phi'^{\pm}
\nonumber\\ && \quad\qquad\qquad\;\;
+\;\frac{\Delta_{\pm}(0)}{2}\; cos\;\rho\; sin\;\rho\;
\left(-1\;+\;2\;\lambda\;\Delta_{\pm}(0)\; sin\;\rho\right)
\left(\phi'^{\pm}\right)^{2} {\bigg ]}\; {\bigg |}_{\rho =\frac{\pi}{2}}
\; ,
\nonumber\\
\label{68}
\ea
where the volume term is the same as the one for the metrical energy
as computed in \cite{freedman}\cite{freedman8}\cite{mezincescu}, and it is
given by

\ba
E^{\pm}_{V}&=&\frac{1}{2}\;\int
d\Omega_{d}\int_{0}^{\frac{\pi}{2}}d\rho\; (cos\;\rho)^{2
\left(\Delta_{\pm}(\varrho )-1\right)}\;(tan \;\rho)^{d-1}\nonumber\\
&\times &
{\bigg [}\;
cos^{2}\rho\left((\partial_{\tau}\phi'^{\pm})^{2}
+(\partial_{\rho}\phi'^{\pm})^{2}+\Delta_{\pm}(\varrho )
\left(\phi'^{\pm}\right)^{2}\right)\nonumber\\ &+&
\frac{1}{tan^{2}\rho}
\left((\partial_{\theta_{1}}\phi'^{\pm})^{2}
+\sum_{i=2}^{d-2}\frac{1}{\prod_{j=1}^{i-1}sin^{2}\theta_{j}}\;
(\partial_{\theta_{i}}\phi'^{\pm})^{2}+\;
\frac{1}{\prod_{j=1}^{d-2}sin^{2}\theta_{j}}\;
(\partial_{\varphi}\phi'^{\pm})^{2}
\right)\; {\bigg ]}\; ,\nonumber\\
\label{69}
\ea
where $\varphi$ and $\theta_{i}\;(1\leq i\leq d-2)$ are spherical
coordinates with $0\leq\varphi <2\pi$ and $0\leq\theta_{i} <\pi$. Note
that the term
$\sum_{i=2}^{d-2}\frac{1}{\prod_{j=1}^{i-1}sin^{2}\theta_{j}}\; 
(\partial_{\theta_{i}}\phi'^{\pm})^{2}$ in the equation above is not
present for $d=3$. We stress the fact that the difference between two
canonical energies, or between a
canonical energy and the metrical one, is just a surface term, as
expected. From
Eqs.(\ref{68'}-\ref{69}), it is straightforward to compute the conditions
for which the canonical energies are positive and finite, and the
procedure is analogous to the one employed in
\cite{freedman}\cite{freedman8}\cite{mezincescu} in the context of the
formulation involving the metrical energy. The volume term
Eq.(\ref{69}) is manifestly positive. Note that for the solution
$\Delta_{+}(\varrho)$ it is also convergent for any value of
$\nu$, whereas for the solution $\Delta_{-}(\varrho)$ it
is convergent only when $\nu$ satisfies the constraint
Eq.(\ref{12000}). On the other hand, the surface terms in 
Eqs.(\ref{68'}-\ref{68}) vanish for the
solution $\Delta_{+}(\varrho)$, but they are divergent for the solution
$\Delta_{-}(\varrho)$, unless we set the constraints Eqs.(\ref{150},
\ref{152},
\ref{154}) which cancel the divergent terms out. Thus, the constraints
which make the canonical energies to be conserved are the same for which
they are also positive and finite.

Once the Cauchy problem has been solved, we can perform the quantization 
in the usual way by regarding the
coefficients $a^{\pm}_{n l\{ m\}}$ and $a^{\pm *}_{n l\{ m\}}$ in
Eq.(\ref{70}) as operators satisfying the commuting relations

\beq
[a^{\pm}_{nl\{ m\}},a^{\pm}_{n'l'\{ m'\}}]=
[a^{\pm *}_{nl\{ m\}},a^{\pm*}_{n'l'\{ m'\}}]=0\; ,
\label{73}
\eeq
\beq
[a^{\pm}_{nl\{ m\}},a^{\pm*}_{n'l'\{
m'\}}]=\delta_{nn'}\;\delta_{ll'}\;\delta_{\{ m\}\{ m'\}}\; .
\label{74}  
\eeq
In particular, for the cases Eqs.(\ref{151}, \ref{153},
\ref{155}) and the masses Eq.(\ref{12000}), which allow for irregular
modes
to propagate in the bulk as well as the regular ones, there exist two
possible consistent quantizations of the scalar field theory on AdS. From
the AdS/CFT correspondence point of view, this means that for such cases
we must have two different boundary CFT's. This topic will be considered
in detail in the next section. 

So far, we have analyzed the case $d\geq 3$. We close this section by
discussing the case $d=2$. The most important
difference between the case $d=2$
and the higher dimensional ones is that for $d=2$ the angular part of the
equation
of motion Eq.(\ref{718'}) has a normalized solution of the form
$\frac{1}{\sqrt{2\pi}}\; e^{-il\varphi}$ (for $0\leq\varphi <2\pi$ and
$l\in
{\bf Z}$) instead of the spherical harmonics $Y_{l\{m\}}(\Omega_{d})$ (see
Eq.(\ref{43})). Note that, unlike the higher dimensional cases, for
$d=2$ the angular number $l$ can also be negative. But for negative $l$ 
the
radial solution Eq.(\ref{43'}) is divergent at the origin. Thus, for $d=2$
we have to make use of $\mid l\mid$ instead of $l$. The remaining
calculations are analogous to the ones performed in the higher
dimensional cases.

So far, we have developed a consistent quantization of the scalar
field theory on
the AdS bulk. We have claimed in Section 2 that the usual AdS/CFT
prescription has no information about the constraints for which
irregular modes propagate in the bulk in the usual quantization along the
lines of \cite{freedman}\cite{freedman8}\cite{mezincescu}. In the
next section, we will consider a generalized AdS/CFT prescription
which, as we will show, gives rise in a natural way to the new constraints
Eqs.(\ref{12000}, \ref{150}, \ref{152}, \ref{154}) for which irregular
modes propagate in the bulk when the canonical
energy is considered instead of the metrical one. In
this way, we will be able to remove all the difficulties that we have 
discussed in Section 2 regarding the usual formulation.

\section{The Generalized AdS/CFT Prescription}

In this section, we present the generalized version of the usual AdS/CFT
prescription. It means that, instead of Eqs.(\ref{9013}, \ref{9015},
\ref{9043}), we
will make use of Eqs.(\ref{9044}-\ref{9044'}). The main goal will be to
show that
this
generalized prescription gives rise, in a natural way, to the constraints
Eqs.(\ref{12000}, \ref{150}, \ref{152}, \ref{154}) for which the irregular
modes propagate in the AdS bulk when the canonical energy is
considered
instead of the metrical one. This provides strong evidence
in support of our
conjectures that the canonical energy is the natural one to be considered
in the context of the AdS/CFT correspondence, and that the generalized
prescription Eqs.(\ref{9044}-\ref{9044'}) is needed in order to map to the
border all the information contained in the bulk. We will show that 
the constraints Eqs.(\ref{12000}, \ref{150}, \ref{152}, \ref{154}) are
the ones for which the
Legendre transformation Eq.(\ref{9045}) interpolates between the conformal
dimensions $\Delta_{+}(\varrho)$ and $\Delta_{-}(\varrho)$, and for
which the divergent local terms in the action cancel out. In this
situation, there is no need to add any counterterms. An interesting result
that we will also find is that there exist particular cases for which the
unitarity bound is reached and the conformal dimension becomes independent
of the mass. This phenomenon has no bulk counterpart.

We work throughout this section in the Euclidean representation of the
$AdS _{d+1}$ in Poincar\'e coordinates described by the
half space $x _{0}>0$, $x _{i} \in {\bf R}$ with metric

\beq
ds^{2}=\frac{1}{x _{0}^{2}} \sum_{\mu=0}^{d} dx^{\mu}dx^{\mu},
\label{3000}
\eeq
where, as in the previous section, we have fixed the
radius of $AdS_{d+1}$ equal to one. We consider the space as foliated by a
family of surfaces $x_{0}=\epsilon$ whose corresponding outward pointing
unit normal vector is

\beq
n_{\mu} = \left(-\epsilon^{-1},{\bf 0}\right)\; ,
\label{3005}
\eeq
from which we find the following trace of the extrinsic curvature

\beq
K=-d\; .
\label{3006}
\eeq
We want to formulate boundary-value problems for the scalar field on the
surfaces $x_{0}=\epsilon$, and then, according to the prescription in
\cite{freedman3}, perform the limit $\epsilon\rightarrow 0$
only at the end of calculations. Performing a Wick rotation, we write the
actions Eqs.(\ref{714}, \ref{720}, \ref{725}) as

\ba
I_{D} &=& \frac{1}{2}\;\int d^{d+1} x \;\sqrt{g}\;
\left[ g^{\mu\nu}\partial_{\mu}\phi\;\partial_{\nu}\phi\; +
\;\left( m^{2}+\;\varrho R\right)\phi^{2}\right]
\; -\;\varrho\int
d^{d}x\;\sqrt{h}\; K_{\epsilon}\;\phi^{2}_{\epsilon}\; ,
\nonumber\\ \label{3001}\\
I_{M,I} &=& I_{D} - \int
d^{d}x\;\sqrt{h}\;\phi_{\epsilon}\;\partial_{n}\phi_{\epsilon}\; ,
\label{3002}\\
I_{M,I\!I} &=& I_{D}{\huge\mid}_{\varrho =0}+\;\lambda\int   
d^{d}x\;\sqrt{h}\;\left(\partial_{n}\phi_{\epsilon}\right)^{2}\; ,
\label{3003}
\ea
where $\phi_{\epsilon}$ is the value of the field at $x_{0}=\epsilon$, and
$\partial_{n}\phi_{\epsilon}$
is its Lie derivative along $n_{\mu}$ given by $\partial_{n}\phi
=n^{\mu}\partial_{\mu}\phi$. Using the equation of motion
Eq.(\ref{718'}), it can easily be shown that the above actions are
stationary under the infinitesimal variation $\phi\rightarrow\phi
+\delta\phi$ for Dirichlet, Type I mixed and Type II mixed boundary
conditions on the scalar field at the surface $x_{0}=\epsilon$,
respectively. Namely

\ba 
I_{D} &\rightarrow &
\delta\phi_{\epsilon}=0\qquad\qquad\;\mbox{(Dirichlet)}\; ,\label{3013}\\
I_{M,I} &\rightarrow & \delta\psi^{I}_{\epsilon}=0\qquad
\qquad\;\mbox{(Type I mixed)}\; ,\label{3014}\\
I_{M,I\! I} &\rightarrow &
\delta\psi^{I\! I}_{\epsilon}=0\qquad\qquad\mbox{(Type II
mixed)}\; ,\label{3015}
\ea
where the fields $\psi^{I}$ and $\psi^{I\! I}$ are defined as in
Eqs.(\ref{723}, \ref{728}).

Integrating by parts and making use of the equation of motion, the actions
Eqs.(\ref{3001}-\ref{3003}) can be written as the following pure-surface
terms

\ba
I_{D} &=& \frac{1}{2}\int
d^{d}x\;\sqrt{h}\;\phi_{\epsilon}\left(\partial_{n}
\phi_{\epsilon}-2\varrho
K_{\epsilon}\phi_{\epsilon}\right)\; ,
\label{3040}\\
I_{M,I} &=& -\frac{1}{2}\int
d^{d}x\;\sqrt{h}\;\phi_{\epsilon}\;\psi_{\epsilon}^{I}\; ,
\label{3041}\\
I_{M,I\! I} &=& \frac{1}{2}\int
d^{d}x\;\sqrt{h}\;\partial_{n}\phi_{\epsilon}\;\psi_{\epsilon}^{I\! I}\; .
\label{3042}
\ea
Note that in the action $I_{D}$, the field
$\partial_{n}\phi_{\epsilon}-2\varrho
K_{\epsilon}\phi_{\epsilon}$ must be written in terms of the boundary data
$\phi_{\epsilon}$. In the case of $I_{M,I}$, $\phi_{\epsilon}$ is to be
written in terms of the data $\psi_{\epsilon}^{I}$. And in the action
$I_{M,I\! I}$, we have to write $\partial_{n}\phi_{\epsilon}$ in terms of
$\psi_{\epsilon}^{I\! I}$. In all calculations, we make use of the
solution of
the equation of motion which is regular at $x_{0}\rightarrow\infty$,
namely \cite{freedman3}

\beq
\phi(x) = \int\frac{d^{d}k}{\left(2\pi\right)^{d}}\;
e^{-i\vec{k}\cdot\vec{x}}\;
x_{0}^{\frac{d}{2}}\;b(\vec{k})\;K_{\nu}(kx_{0}),
\label{3043}
\eeq
where $\vec{x}=(x^{1},...,x^{d})$, $k = \mid\vec{k}\mid$,
$K_{\nu}$ is the modified Bessel function, and $\nu$ is given by
Eq.(\ref{9006'''}). From the equation above, we also get

\beq
\partial_{n}\phi(x) = -\int\frac{d^{d}k}{\left(2\pi\right)^{d}}\;
e^{-i\vec{k}\cdot\vec{x}}\;
x_{0}^{\frac{d}{2}}\;b(\vec{k})\;\left[\left(\frac{d}{2} + \nu
\right)K_{\nu}(kx_{0}) - kx_{0}K_{\nu + 1}(kx_{0})\right].
\label{3044}
\eeq

In the following sub-sections, we will consider each boundary condition
separately.

\subsection{Dirichlet Boundary Condition}

Here we will consider the generalized AdS/CFT prescription
Eqs.(\ref{9044}-\ref{9044'}) in the
particular case of Dirichlet boundary conditions, and the key result will
be to reproduce, in a natural way, the constraints
Eqs.(\ref{12000}, \ref{150}) obtained in the
bulk.

Let $\phi_{\epsilon}(\vec{k})$ be the
Fourier transform
of the Dirichlet boundary value of the field
$\phi_{\epsilon}(\vec{x})$. From
Eq.(\ref{3043}) we find

\beq
b(\vec{k})=\frac{\epsilon^{-\frac{d}{2}}\;\phi_{\epsilon}(\vec{k})}
{K_{\nu}(k\epsilon)}\; ,
\label{3045}
\eeq
and, inserting this into Eq.(\ref{3044}), we write
$\partial_{n}\phi_{\epsilon}$ in terms of the boundary data
$\phi_{\epsilon}$ as

\beq
\partial_{n}\phi_{\epsilon}(\vec{x})=-\int
d^{d}y\;\phi_{\epsilon}(\vec{y})\int\frac{d^{d}k}{\left(
2\pi\right)^{d}}\;e^{-i\vec{k}\cdot\left(
\vec{x}-\vec{y}\right)}\;\left[\frac{d}{2} + \nu -
k\epsilon\;\frac{K_{\nu+1}(k\epsilon)}{K_{\nu}(k\epsilon)}\right].
\label{3046}
\eeq
Using this together with Eq.(\ref{3006}), we can write the action
Eq.(\ref{3040}) as

\ba
I_{D}\left[\phi_{\epsilon}\right] &=& -\frac{1}{2}\int d^{d}x
\; d^{d}y\;\sqrt{h}\;\phi_{\epsilon}(\vec{x})
\;\phi_{\epsilon}(\vec{y})\nonumber\\
&&\qquad\times\int\frac{d^{d}k}{\left(
2\pi\right)^{d}}\;e^{-i\vec{k}\cdot\left(
\vec{x}-\vec{y}\right)}\;\left[\frac{d}{2} + \nu -2\varrho\; d-
k\epsilon\;\frac{K_{\nu+1}(k\epsilon)}{K_{\nu}(k\epsilon)}\right].  
\label{3047}
\ea
This is only one of the two functionals which contain the information
about the boundary CFT's. The another one is obtained by performing the
Legendre transformation Eq.(\ref{9045}), with the identifications
$A_{\epsilon}(\vec{x})\equiv\phi_{\epsilon}(\vec{x})$,
${\tilde A}_{\epsilon}(\vec{x})\equiv{\tilde\phi}_{\epsilon}(\vec{x})$, on
the
action above. This is written as

\beq
{\cal J}_{D}\left[\phi_{\epsilon},{\tilde \phi}_{\epsilon}\right]\;=\;
I_{D}\left[\phi_{\epsilon}\right]\; -\int
d^{d}x\;\sqrt{h}\;\phi_{\epsilon}(\vec{x})\;
{\tilde\phi}_{\epsilon}(\vec{x})\; ,
\label{3048'}
\eeq
which, after setting $\frac{\partial {\cal
J}_{D}}{\partial\phi_{\epsilon}}=0$, gives rise to the
identity \footnote{Note that the global minus sign
at the r.h.s of Eq.(\ref{3047'}) is just conventional and has no physical
meaning, because it cancels out on computing the transformed functional.}

\beq
\phi_{\epsilon}=-\frac{1}{\frac{d}{2} + \nu -2\varrho\; d-
k\epsilon\;\frac{K_{\nu+1}(k\epsilon)}{K_{\nu}(k\epsilon)}} 
\;{\tilde\phi}_{\epsilon}\; .
\label{3047'}
\eeq
Then, the Legendre transform of $I_{D}\left[\phi_{\epsilon}\right]$ reads

\beq
{\tilde I}_{D}\left[{\tilde \phi}_{\epsilon}\right]=\frac{1}{2}\int
d^{d}x
\; d^{d}y\;\sqrt{h}\;{\tilde\phi}_{\epsilon}(\vec{x})
\;{\tilde\phi}_{\epsilon}(\vec{y})\int\frac{d^{d}k}{\left(
2\pi\right)^{d}}\;e^{-i\vec{k}\cdot\left(
\vec{x}-\vec{y}\right)}\;\frac{1}{\frac{d}{2} + \nu -2\varrho\; d-
k\epsilon\;\frac{K_{\nu+1}(k\epsilon)}{K_{\nu}(k\epsilon)}}\; .
\label{3048}
\eeq
Both functionals $I_{D}$ and ${\tilde I}_{D}$ are needed in order to map
to the boundary all the information contained in the bulk. Each one of the
fields $\phi_{\epsilon}$ and ${\tilde\phi}_{\epsilon}$ will couple, after
performing the limit $\epsilon\rightarrow 0$ in a proper way and through
the prescriptions Eqs.(\ref{9044}, \ref{9044'}), to the corresponding
boundary conformal operator. The transformed functional
${\tilde I}_{D}\left[{\tilde \phi}_{\epsilon}\right]$ is to be contrasted
with the usual one Eq.(\ref{9014}), which contains only a non-local
term. In the transformed functional ${\tilde I}_{D}\left[{\tilde
\phi}_{\epsilon}\right]$ we
will find, after
performing a series expansion in powers of $\epsilon$, both local and
non-local terms, just as it happens to the original functional
Eq.(\ref{3047}). An important result will be that the conformal dimension
$\Delta_{-}(\varrho)$ arises precisely when the divergent local terms
cancel out, showing that they contain information about the
leading non-local term, and need to be taken into account. And the key
result that we will
show is that such cancellation of divergent terms arises, precisely, for
the constraints Eqs.(\ref{12000}, \ref{150}) obtained in the bulk. Note
that, unlike what happens to
the usual transformed
functional Eq.(\ref{9014}), the functional ${\tilde I}_{D}\left[{\tilde
\phi}_{\epsilon}\right]$ contains
no coefficient to be fixed `by hand'. Besides, the particular case of
$\nu =0$, which as explained in Section 2 gives rise to some difficulties
in the usual formulation, will be successfully analyzed in this generalized
context, through the simple procedure of performing a series
expansion of the ratio
$k\epsilon\;\frac{K_{1}(k\epsilon)}{K_{0}(k\epsilon)}$
in powers of $\epsilon$.

We begin by concentrating on the case of $\nu$ not integer. Note
that, by making use of the series expansion

\ba
k\epsilon\;\frac{K_{\nu
+1}(k\epsilon)}{K_{\nu}(k\epsilon)} &=& 2\nu\;
{\bigg [}\;\left(
1\; -\;\frac{1}{4\nu(1-\nu)}\; (k\epsilon)^{2}\;
 + \; O(\epsilon^{4})\right)\nonumber\\  && \qquad +\;
\left(2^{-2\nu}\;\frac{\Gamma (1-\nu)}{\Gamma
(1+\nu)}\; (k\epsilon)^{2\nu} \; +\; O(\epsilon^{2\nu
+2})\right)\; {\bigg ]}\; ,
\label{3049}
\ea
we can write Eq.(\ref{3047'}) as

\beq
\phi_{\epsilon}=-\left[\Delta_{-}(\varrho)\; -\;2\varrho\; d
\; +\;\frac{1}{2(1-\nu)}\;(k\epsilon)^{2}\; -\; 2^{1-2\nu}
\;\frac{\Gamma (1-\nu)}{\Gamma
(\nu)}\; (k\epsilon)^{2\nu}\; +\;\cdots\right]^{-1}\; {\tilde 
\phi}_{\epsilon}\; , 
\label{3050}
\eeq
where the dots stand for higher order terms in
$\epsilon$. The functionals $I_{D}$ and ${\tilde I}_{D}$ read

\ba
I_{D}[\phi_{\epsilon}] &=& -\frac{1}{2}\int d^{d}x
\; d^{d}y\;\phi_{\epsilon}(\vec{x})
\;\phi_{\epsilon}(\vec{y})\;\epsilon^{-d}\nonumber\\
&&\qquad\times\int\frac{d^{d}k}{\left(
2\pi\right)^{d}}\;e^{-i\vec{k}\cdot\left(
\vec{x}-\vec{y}\right)}\;{\bigg [}\;\Delta_{-}(\varrho)\; -\;2\varrho\; d
\; +\;\frac{1}{2(1-\nu)}\;(k\epsilon)^{2}\nonumber\\
&&\qquad\qquad\qquad\qquad\qquad\qquad - 2^{1-2\nu}
\;\frac{\Gamma (1-\nu)}{\Gamma
(\nu)}\; (k\epsilon)^{2\nu}\; +\;\cdots\; {\bigg ]}\; ,
\label{3080}
\ea

\ba
{\tilde I}_{D}[{\tilde\phi}_{\epsilon}] &=&\frac{1}{2}\int d^{d}x
\; d^{d}y\;{\tilde\phi}_{\epsilon}(\vec{x})
\;{\tilde\phi}_{\epsilon}(\vec{y})\;\epsilon^{-d}\nonumber\\
&&\qquad\times\int\frac{d^{d}k}{\left(
2\pi\right)^{d}}\;e^{-i\vec{k}\cdot\left(
\vec{x}-\vec{y}\right)}\;{\bigg [}\;\Delta_{-}(\varrho)\; -\;2\varrho\; d
\; +\;\frac{1}{2(1-\nu)}\;(k\epsilon)^{2}\nonumber\\
&&\qquad\qquad\qquad\qquad\qquad\qquad - 2^{1-2\nu}
\;\frac{\Gamma (1-\nu)}{\Gamma
(\nu)}\; (k\epsilon)^{2\nu}\; +\;\cdots\; {\bigg ]}^{-1}\; .
\label{3081}
\ea

We consider first the situation in which the constraint Eq.(\ref{150}) is
not satisfied, i.e. we have $\varrho\not= \varrho^{\pm}_{D}$, with
$\varrho^{\pm}_{D}$ given by Eq.(\ref{151}). Then Eq.(\ref{3050}) reads

\beq
\phi_{\epsilon} =
-\left[\frac{1}{\Delta_{-}(\varrho)\; -\;2\varrho\; d}+\cdots\right]
\; {\tilde \phi}_{\epsilon}\qquad (\varrho\not= \varrho^{\pm}_{D})\; ,
\label{3051}
\eeq
and the fields $\phi_{\epsilon}$ and ${\tilde \phi}_{\epsilon}$ have
the same asymptotic behavior. This means that, in this particular case,
the Legendre transformation interpolates between boundary operators of the
same conformal dimension (which will happen to be $\Delta_{+}(\varrho)$,
as expected). This is a novel situation, since the usual Legendre
transform prescription Eq.(\ref{9013}) only interpolates between different
conformal dimensions. However, our result is in agreement with the fact
that, when the constraint Eq.(\ref{150}) is not satisfied, only
regular modes propagate in the bulk, thus leading to expect to find
only one conformal dimension at the boundary (namely
$\Delta_{+}(\varrho)$). In this situation, the divergent local terms in
Eqs.(\ref{3080}, \ref{3081}) do not cancel out. They encode the
information, through Eq.(\ref{3051}), that in this case the Legendre
transformation interpolates between boundary operators of the same
conformal
dimension. Integrating over $\vec{k}$ in Eqs.(\ref{3080}, \ref{3081}) we
get

\ba
I^{\varrho\not= \varrho^{\pm}_{D}}_{D}[\phi_{\epsilon}] &=&
\mbox{divergent local terms}
\nonumber\\
 && - \frac{\nu (\varrho)}{\pi^{\frac{d}{2}}}\;
\frac{\Gamma(\Delta_{+}(\varrho))}{\Gamma(\nu (\varrho))}
\; \int d^{d}x \;
d^{d}y\;\phi_{\epsilon}(\vec{x})
\;\phi_{\epsilon}(\vec{y})\;\frac{\epsilon^{-2\Delta_{-}(\varrho)}}{|\;
\vec{x}-\vec{y}\;|^{2\Delta_{+}(\varrho)}}\;+\;\cdots,
\nonumber\\
\label{3090} \\
{\tilde I}^{\varrho\not= \varrho^{\pm}_{D}}_{D}[{\tilde \phi}_{\epsilon}]
&=& \mbox{divergent local terms}
\nonumber\\
 && - \frac{\nu (\varrho)}{\pi^{\frac{d}{2}}}\;
\frac{1}{\left[\Delta_{-}(\varrho)-2\varrho\; d\right]^{2}}
\;\frac{\Gamma(\Delta_{+}(\varrho))}{\Gamma(\nu (\varrho))}
\; \int d^{d}x \;
d^{d}y\;{\tilde \phi}_{\epsilon}(\vec{x})
\;{\tilde 
\phi}_{\epsilon}(\vec{y})\;\frac{\epsilon^{-2\Delta_{-}(\varrho)}}{|\;
\vec{x}-\vec{y}\;|^{2\Delta_{+}(\varrho)}}
\nonumber\\ && +\cdots,
\label{3091}
\ea
where the dots stand for higher order terms in $\epsilon$. The non-local
term in Eq.(\ref{3090}) is the same as the one analyzed in
\cite{freedman3}\cite{viswa1}, with an effective mass
Eq.(\ref{9016'}). In this case, the generalized AdS/CFT prescription
Eq.(\ref{9044}) reduces to the usual Dirichlet prescription
Eq.(\ref{9043}). The
non-local term in Eq.(\ref{3091}) differs from the one in
Eq.(\ref{3090}) only by a normalization factor. Taking the limits

\beq
\lim _{\epsilon\rightarrow
0}\epsilon^{-\Delta_{-}(\varrho)}\;\phi_{\epsilon}(\vec{x})
= \phi_{0}(\vec{x}),
\label{3092}
\eeq

\beq
\lim _{\epsilon\rightarrow
0}\epsilon^{-\Delta_{-}(\varrho)}\;{\tilde \phi}_{\epsilon}(\vec{x})
= {\tilde \phi}_{0}(\vec{x}),  
\label{3093}
\eeq
to go to the border, and making use of the AdS/CFT prescription

\beq
\exp\left( -I_{D}[\phi_{0}]\right) \equiv \left<\exp\left(\int
d^{d}x \;   
{\cal O}_{D}(\vec{x}) \; \phi_{0}(\vec{x})\right)\right>,
\label{3094}
\eeq

\beq
\exp\left( -{\tilde I}_{D}[{\tilde\phi}_{0}]\right) \equiv
\left<\exp\left(\int d^{d}x \;
{\tilde {\cal O}}_{D}(\vec{x}) \; {\tilde
\phi}_{0}(\vec{x})\right)\right>,
\label{3095}
\eeq
we find the following boundary two-point functions
\ba
\left<{\cal O}^{\varrho\not= \varrho^{\pm}_{D}}_{D}(\vec{x})\; {\cal
O}^{\varrho\not= \varrho^{\pm}_{D}}_{D}(\vec{y})\right> &=&
\frac{2\nu(\varrho)}{\pi^{\frac{d}{2}}}\;
\frac{\Gamma(\Delta_{+}(\varrho))}{\Gamma(\nu(\varrho))}\;\frac{1}{|\;
\vec{x}-\vec{y}\;|^{2\Delta_{+}(\varrho)}}
\nonumber\\ \nonumber\\ &+& \;\mbox{divergences}\; ,
\label{3096}
\\
\nonumber
\\
\left<{\tilde {\cal
O}}^{\varrho\not= \varrho^{\pm}_{D}}_{D}(\vec{x})\; {\tilde {\cal
O}}^{\varrho\not= \varrho^{\pm}_{D}}_{D}(\vec{y})\right> &=&
\frac{2\nu(\varrho)}{\pi^{\frac{d}{2}}}\;
\frac{1}{\left[\Delta_{-}(\varrho)-2\varrho\; d\right]^{2}}
\;\frac{\Gamma(\Delta_{+}(\varrho))}{\Gamma(\nu(\varrho))}\;\frac{1}{|\;
\vec{x}-\vec{y}\;|^{2\Delta_{+}(\varrho)}}\nonumber\\ \nonumber\\ 
&+& \;\mbox{divergences}\; ,
\label{3097}
\ea
where the divergences are due to the fact that the functionals
Eqs.(\ref{3090}, \ref{3091}) contain divergent local terms. In this
situation, both conformal operators ${\cal O}^{\varrho\not=
\varrho^{\pm}_{D}}_{D}$ and ${\tilde {\cal O}}^{\varrho\not= 
\varrho^{\pm}_{D}}_{D}$ have the same conformal dimension
$\Delta_{+}(\varrho)$, as expected from the analysis in Section 3 which
establishes that for $\varrho\not= \varrho^{\pm}_{D}$ only regular modes 
propagate in the bulk when the canonical energy is taken into
account. From Eqs.(\ref{3092}, \ref{3093}), we conclude that near the 
border the classical modes, which encode the information about the
boundary CFT, behave as $x_{0}^{\Delta_{-}(\varrho)}$, also as
expected from the results in Section 3. The Legendre transformation
interpolates between conformal operators of
the same conformal dimension. We remark that this information is
encoded in the fact that the divergent local terms of the functionals
$I^{\varrho\not= \varrho^{\pm}_{D}}_{D}[\phi_{\epsilon}]$ and
${\tilde I}^{\varrho\not=
\varrho^{\pm}_{D}}_{D}[{\tilde\phi}_{\epsilon}]$ do not cancel out. In
this case, the only difference between the boundary two-point functions
Eqs.(\ref{3096}, \ref{3097}) is just a normalization factor.

Now we consider the case in which the constraint Eq.(\ref{150}) is
satisfied, i.e. we have $\varrho = \varrho^{\pm}_{D}$. In this situation 
we expect, from the analysis in Section 3, to find two different boundary
CFT's, namely the ones of conformal dimensions $\Delta_{+}(\varrho)$ and
$\Delta_{-}(\varrho)$. Unlike the case of $\varrho
\not= \varrho^{\pm}_{D}$, in the situation $\varrho =\varrho^{\pm}_{D}$ we
must analyze two different
cases, namely the ones in which the constraint Eq.(\ref{12000}) is,
or not, also satisfied. We begin by concentrating on the situation in
which the
constraint Eq.(\ref{12000}) is also satisfied (with the particular
case $\nu =0$ naturally excluded, since we are considering here that
$\nu$ is not an integer).

Let us first analyze the original functional 
$I_{D}[\phi_{\epsilon}]$. Integrating
over $\vec{k}$ in Eq.(\ref{3080}) we find

\ba
I^{\varrho = \varrho^{\pm}_{D}\; ;\;\nu <1}_{D}[\phi_{\epsilon}]
&=&
- \frac{\nu(\varrho)}{\pi^{\frac{d}{2}}}\;
\frac{\Gamma(\Delta_{+}(\varrho))}{\Gamma(\nu (\varrho))}
\; \int d^{d}x \;
d^{d}y\;\phi_{\epsilon}(\vec{x})
\;\phi_{\epsilon}(\vec{y})\;\frac{\epsilon^{-2\Delta_{-}(\varrho)}}{|\;
\vec{x}-\vec{y}\;|^{2\Delta_{+}(\varrho)}}\;+\;\cdots,
\nonumber\\
\label{3100}
\ea
where the dots stand for higher order terms in $\epsilon$. Notice that
the only, and crucial difference between Eqs.(\ref{3090},
\ref{3100}), is
that now the divergent local terms have cancelled out. This fact will
acquire great relevance when we consider the transformed functional
${\tilde I}_{D}[{\tilde\phi}_{\epsilon}]$. 

Taking the limit Eq.(\ref{3092}), we find the two-point function

\ba
\left<{\cal O}^{\varrho = \varrho^{\pm}_{D}\; ;\;\nu 
<1}_{D}(\vec{x})\; {\cal   
O}^{\varrho = \varrho^{\pm}_{D}\; ;\;\nu <1}_{D}(\vec{y})\right> &=&
\frac{2\nu(\varrho)}{\pi^{\frac{d}{2}}}\;
\frac{\Gamma(\Delta_{+}(\varrho))}{\Gamma(\nu(\varrho))}\;\frac{1}{|\;
\vec{x}-\vec{y}\;|^{2\Delta_{+}(\varrho)}}\; ,
\label{3105}
\ea
which contains the same non-local term as the two-point function
Eq.(\ref{3096}), as expected from the fact that the functionals
Eqs.(\ref{3090}, \ref{3100}) share the same leading non-local
term, but that, unlike Eq.(\ref{3096}), does not contain any
divergences. Thus, when the constraint Eq.(\ref{150}) is satisfied, the
divergences disappear. The
conformal operator ${\cal O}^{\varrho = \varrho^{\pm}_{D}\; ;\;\nu
<1}_{D}$ has
conformal dimension $\Delta_{+}(\varrho)$, and the classical
bulk modes to which it couples behave close to the boundary as
$x_{0}^{\Delta_{-}(\varrho)}$. 

So far, we have been able to reproduce only the conformal dimension
$\Delta_{+}(\varrho)$. From the analysis in Section 3, we expect to find
also the conformal dimension $\Delta_{-}(\varrho)$. Let us show that it 
arises on considering the transformed functional
${\tilde I}_{D}[{\tilde\phi}_{\epsilon}]$. Making use of Eqs.(\ref{12000},
\ref{150}), we note that Eq.(\ref{3050}) can be written as

\beq
\phi_{\epsilon} = \left[2^{2\nu -1}\;\frac{\Gamma (\nu)}{\Gamma
(1-\nu)}\; (k\epsilon)^{-2\nu}+\cdots\right]\; {\tilde \phi}_{\epsilon} 
\; ,
\label{3052}
\eeq
where the dots stand for higher order terms in $\epsilon$. In this
case, the fields $\phi_{\epsilon}$ and ${\tilde \phi}_{\epsilon}$
have different asymptotic behaviors, and thus the Legendre transformation 
Eq.(\ref{3048'}) interpolates between two different conformal
dimensions. Since, as we have shown, $\phi_{\epsilon}$ corresponds to the
conformal dimension $\Delta_{+}(\varrho)$, it is clear from
Eq.(\ref{3052}) that ${\tilde \phi}_{\epsilon}$ corresponds to the missing
conformal dimension
$\Delta_{-}(\varrho)$, as expected. The information that the Legendre
transformation interpolates between two different conformal dimensions is
encoded in
the fact that, in this case, the divergent local terms in Eqs.(\ref{3080},
\ref{3081}) cancel out. The constraints Eqs.(\ref{12000},
\ref{150}) arise, in this generalized AdS/CFT prescription, in a natural
way, as anticipated. Integrating over $\vec{k}$ in Eq.(\ref{3081}) we get

\ba
{\tilde I}^{\varrho =
\varrho^{\pm}_{D}\; ;\;\nu <1}_{D}[{\tilde
\phi}_{\epsilon}]
&=& 
-\frac{1}{4\pi^{\frac{d}{2}}}
\;\frac{\Gamma(\Delta_{-}(\varrho))}{\Gamma(1-\nu(\varrho))}\int d^{d}x \;
d^{d}y\;{\tilde\phi}_{\epsilon}(\vec{x})
\;{\tilde\phi}_{\epsilon}(\vec{y})
\;\frac{\epsilon^{-2\Delta_{+}(\varrho)}}{|\;
\vec{x}-\vec{y}\;|^{2\Delta_{-}(\varrho)}}
\nonumber\\ &&\quad +\;\cdots,
\label{3101}
\ea
where the dots stand for higher order terms in $\epsilon$. Notice that
the fact that the divergent local terms have cancelled out has 
given rise to the presence, in the transformed functional
${\tilde I}_{D}[{\tilde\phi}_{\epsilon}]$, of a new (with respect to the
one in
Eq.(\ref{3091})) leading non-local term, which, as we will show,
corresponds
to a boundary CFT of conformal dimension $\Delta_{-}(\varrho)$, as
expected.

Taking the limit

\beq
\lim _{\epsilon\rightarrow
0}\epsilon^{-\Delta_{+}(\varrho)}\;{\tilde \phi}_{\epsilon}(\vec{x})
= {\tilde \phi}_{0}(\vec{x}),
\label{3120}
\eeq
we find the following two-point function

\ba
\left<{\tilde {\cal O}}^{\varrho =
\varrho^{\pm}_{D}\; ;\;\nu <1}_{D}(\vec{x})\; {\tilde {\cal  
O}}^{\varrho = \varrho^{\pm}_{D}
\; ;\;\nu <1}_{D}(\vec{y})\right> &=&
\frac{1}{2\pi^{\frac{d}{2}}}\;
\frac{\Gamma(\Delta_{-}(\varrho))}{\Gamma(1-\nu(\varrho))}\;\frac{1}{|\;
\vec{x}-\vec{y}\;|^{2\Delta_{-}(\varrho)}}\; ,
\label{3121}
\ea
and the conformal operator ${\tilde {\cal O}}^{\varrho =
\varrho^{\pm}_{D}\; ;\;\nu <1}_{D}$ has conformal dimension
$\Delta_{-}(\varrho)$, as expected. Besides, in this situation we note
from Eq.(\ref{3120}) that the classical modes which encode the information
about the boundary CFT behave asymptotically as 
$x_{0}^{\Delta_{+}(\varrho)}$. In this way, we have been able, through the
generalized AdS/CFT prescription Eqs.(\ref{9044}-\ref{9044'}), to
reproduce in a natural way the constraints Eqs.(\ref{12000},
\ref{150}) for which irregular modes propagate in the bulk for
Dirichlet boundary conditions and when the
canonical energy, instead of the metrical one, is taken into
account. As anticipated, we
have also shown that irregular modes arise when the divergent
local terms of the action cancel out, and that such divergent terms encode
the information about the leading non-local terms, in the sense that when
they cancel out, and only then, the Legendre transformation
Eq.(\ref{9045}) interpolates
between different conformal dimensions. In this situation, the
addition of counterterms to the on-shell action is not required. Note
that our prescription has left no coefficient to be fixed `by hand'.

We still need to analyze the case in which the constraint
Eq.(\ref{150}) is satisfied, but the constraint Eq.(\ref{12000}) is not,
i.e. we  have $\varrho = \varrho^{\pm}_{D}$ and $\nu>1$. This is
a very particular situation for which, as we will show, the unitarity
bound $(d-2)/2$ arises. 
We begin by considering the original functional
$I_{D}[\phi_{\epsilon}]$. Integration over $\vec{k}$ in
Eq.(\ref{3080}) yields

\ba
I^{\varrho = \varrho^{\pm}_{D}\; ;\;\nu >1}_{D}[\phi_{\epsilon}]
&=&
\mbox{divergent local terms}
\nonumber\\
 && - \frac{\nu (\varrho)}{\pi^{\frac{d}{2}}}\;
\frac{\Gamma(\Delta_{+}(\varrho))}{\Gamma(\nu (\varrho))}
\; \int d^{d}x \;
d^{d}y\;\phi_{\epsilon}(\vec{x})
\;\phi_{\epsilon}(\vec{y})\;\frac{\epsilon^{-2\Delta_{-}(\varrho)}}{|\;
\vec{x}-\vec{y}\;|^{2\Delta_{+}(\varrho)}}\;+\;\cdots,
\nonumber\\
\label{3130}
\ea
and then, unlike the case $\nu<1$ (see Eq.(\ref{3100})), in this
situation the original functional contains divergent local terms,
because the leading order in Eq.(\ref{3080}) turns out to be
$O\left(\epsilon^{-d+2}\right)$. Taking
the limit Eq.(\ref{3092}), we find the two-point function
  
\ba
\left<{\cal O}^{\varrho = \varrho^{\pm}_{D}\; ;\;\nu 
>1}_{D}(\vec{x})\; {\cal
O}^{\varrho = \varrho^{\pm}_{D}\; ;\;\nu >1}_{D}(\vec{y})\right> &=&
\frac{2\nu(\varrho)}{\pi^{\frac{d}{2}}}\;
\frac{\Gamma(\Delta_{+}(\varrho))}{\Gamma(\nu(\varrho))}\;\frac{1}{|\;
\vec{x}-\vec{y}\;|^{2\Delta_{+}(\varrho)}}\nonumber\\ \nonumber\\
&+&\; \mbox{divergences}\; ,
\label{3131}
\ea
which, unlike the two-point function corresponding to the case
$\nu<1$ (see Eq.(\ref{3105})), also contains divergences. The
conformal operator ${\cal O}^{\varrho = \varrho^{\pm}_{D}\; ;\;\nu
>1}_{D}$ has
conformal dimension $\Delta_{+}(\varrho)$, and the classical
bulk modes to which it couples behave close to the boundary as
$x_{0}^{\Delta_{-}(\varrho)}$.

Now we consider the transformed functional ${\tilde
I}_{D}[{\tilde\phi}_{\epsilon}]$. Integrating over $\vec{k}$ in
Eq.(\ref{3081}), we get for $d\geq 3$

\beq
{\tilde I}^{\varrho = \varrho^{\pm}_{D}\; ;\;\nu
>1}_{D}[{\tilde\phi}_{\epsilon}]
=-\left(\nu(\varrho)-1\right)
\frac{\Gamma(\frac{d-2}{2})}{4\pi^{\frac{d}{2}}}\int d^{d}x \;
d^{d}y\;{\tilde\phi}_{\epsilon}(\vec{x})
\;{\tilde\phi}_{\epsilon}(\vec{y})\;
\frac{\epsilon^{-d-2}}{|\;
\vec{x}-\vec{y}\;|^{2\frac{d-2}{2}}}\; +\;\cdots ,
\label{3132}
\eeq
which, unlike the original functional Eq.(\ref{3130}), does not contain
any divergent terms. Taking the limit

\beq
\lim _{\epsilon\rightarrow
0}\epsilon^{-\frac{d+2}{2}}{\tilde\phi}_{\epsilon}(\vec{x})
= {\tilde\phi}_{0}(\vec{x}),  
\label{3133}
\eeq
we find the boundary two-point function

\beq
\left<{\tilde {\cal O}}_{D}^{\varrho =
\varrho^{\pm}_{D}\; ;\;\nu
>1}(\vec{x})\; {\tilde{\cal
O}}_{D}^{\varrho = \varrho^{\pm}_{D}\; ;\;\nu
>1}(\vec{y})\right>=
\left(\nu(\varrho)-1\right)\;
\frac{\Gamma\left( \frac{d-2}{2}\right)}{2\pi^{\frac{d}{2}}}\;
\frac{1}{|\;
\vec{x}-\vec{y}\;|^{2\frac{d-2}{2}}}\; ,
\label{3134}
\eeq
which does not contain any divergences. The operator ${\tilde {\cal
O}}_{D}^{\varrho =\varrho^{\pm}_{D}\; ;\;\nu >1}$ has conformal dimension
$\frac{d-2}{2}$, which is precisely the unitarity bound. This is a very
particular case which does not have any corresponding bulk
counterpart. Note from Eq.(\ref{3133}) that the classical backgrounds
which couple to ${\tilde {\cal
O}}_{D}^{\varrho =\varrho^{\pm}_{D}\; ;\;\nu >1}$ behave close to the
boundary as $x_{0}^{\frac{d+2}{2}}$, and, as noted from Eq.(\ref{52}),
such modes do not exist in the bulk. From the bulk point of view, the
irregular modes
propagate for $0\leq\nu <1$. For $\nu\rightarrow 1$, when the
unitarity bound is reached, they just stop propagating. Then, for
$\nu >1$, only regular modes propagate in the bulk. On the other
hand, from the boundary point of view, the conformal dimension
$\Delta_{-}(\varrho)$ is present for $0\leq\nu <1$. For
$\nu\rightarrow 1$, when the unitarity bound is reached, it does not
disappear. What we find is that the unitarity bound gets saturated, and
then, for $\nu >1$,
the conformal dimension $\Delta_{-}(\varrho)$ becomes independent of the
effective mass Eq.(\ref{9016'}). It just stays as $(d-2)/2$, no matter how
much the effective mass grows up. This is a puzzling phenomenon, which has 
no bulk counterpart, and deserves further studies. We stress the fact
that the conformal dimension $(d-2)/2$ arises only for $d\geq 3$.

So far, we have considered the case of $\nu$ not integer, and now we
analyze the case of $\nu$ integer. In such situation, the series
expansion of
the modified Bessel function contains extra logarithmic terms which must
be
considered. For the case of $\nu$ integer but non-zero, it is possible to
show, following a procedure analogous to the
one in \cite{our2}, that, after integration over $\vec{k}$, the non-local
terms result identical to the ones of the case of $\nu$ not integer, thus
giving rise to the same boundary two-point functions. On the other hand,
the logarithmic divergent local terms, which are not present for $\nu$
not integer, are related to conformal
anomalies
\cite{skenderis}\cite{nojiri}\cite{odintsov2}\cite{petkou}\cite{haro}.
We will not develop this topic further.

We finish the analysis of the Dirichlet boundary condition by considering
in detail the case $\nu =0$. We will show that, in this situation,
we reproduce the expected results, namely that the Legendre transformation
interpolates between boundary operators of the same conformal dimension
$\frac{d}{2}$. As expected, in addition to the usual divergent local 
terms, we will find logarithmic divergences. We will show that, when 
the constraint Eq.(\ref{150}) is satisfied, the usual divergent
local terms cancel out, and only the logarithmic divergences
survive. Another interesting result that we will show regards the
asymptotic behavior of the classical modes which couple to the
boundary operator. In general, the case of $\nu =0$ involves a novel
situation, in which such classical modes present a logarithmic
behavior close to the boundary \cite{freedman3}. A new result that we will
show is that, when the constraint Eq.(\ref{150}) is satisfied, the
Legendre transformation interpolates between conformal operators of the
same
conformal dimension $\frac{d}{2}$, but corresponding to classical
bulk modes which differ in their asymptotic behaviors. One of such
behaviors is logarithmic,
as expected. However, the another one is not, thus resembling the bulk
modes which arise for $\nu$
not integer.

Making use of standard identities for the modified Bessel functions, we
find

\beq
k\epsilon\;\frac{K_{1}(k\epsilon)}{K_{0}(k\epsilon)}=
-\frac{1}{ln\;\epsilon}\;\left[\; 1\; +\;
\frac{ln\; 2\; -\gamma}{ln\;\epsilon}\; -\;
\frac{ln\; k}{ln\;\epsilon}\; +\; O
\left(\epsilon^{2}\; ln\;\epsilon\right)\right]\; ,
\label{3300}
\eeq
where $\gamma$ is the Euler constant. Then, the functionals
Eqs.(\ref{3047}, \ref{3048}) can be written as

\ba
I^{\nu =0}_{D}[\phi_{\epsilon}] &=& -\frac{1}{2}\int d^{d}x
\; d^{d}y\;\phi_{\epsilon}(\vec{x})
\;\phi_{\epsilon}(\vec{y})\;\epsilon^{-d}\nonumber\\
&&\qquad\times\int\frac{d^{d}k}{\left(
2\pi\right)^{d}}\;e^{-i\vec{k}\cdot\left(
\vec{x}-\vec{y}\right)}\;{\bigg
[}\;\frac{d}{2}\; -
2\varrho\; d\; +\; \frac{1}{ln\;\epsilon}\; +\; \frac{ln\; 2\; -\gamma}
{ln^{2}\epsilon}\; -\; \frac{ln\; k}{ln^{2}\epsilon}\nonumber\\ &&
\qquad\qquad\qquad\qquad\qquad\qquad +\; \cdots {\bigg ]}\; ,
\label{3301}
\ea

\ba
{\tilde I}^{\nu =0}_{D}[{\tilde\phi}_{\epsilon}] &=& \frac{1}{2}\int
d^{d}x
\; d^{d}y\;{\tilde\phi}_{\epsilon}(\vec{x})
\;{\tilde\phi}_{\epsilon}(\vec{y})\;\epsilon^{-d}\nonumber\\
&&\qquad\times\int\frac{d^{d}k}{\left(
2\pi\right)^{d}}\;e^{-i\vec{k}\cdot\left(
\vec{x}-\vec{y}\right)}\;{\bigg
[}\;\frac{d}{2}\; -
2\varrho\; d\; +\; \frac{1}{ln\;\epsilon}\; +\; \frac{ln\; 2\; -\gamma}
{ln^{2}\epsilon}\; -\; \frac{ln\; k}{ln^{2}\epsilon}\nonumber\\ &&
\qquad\qquad\qquad\qquad\qquad\qquad +\; \cdots {\bigg ]}^{-1}\; ,
\label{3302}
\ea
where the dots stand for higher order terms in $\epsilon$. Let us first
consider the original functional $I^{\nu =0}_{D}[\phi_{\epsilon}]$. After
integration over $\vec{k}$, the leading non-local term reads

\beq
I^{non-local}_{D}[\phi_{\epsilon}] = 
-\frac{\Gamma\left(
\frac{d}{2}
\right)}{4\pi^{\frac{d}{2}}}\; \int d^{d}x \;
d^{d}y\;\phi_{\epsilon}(\vec{x})
\;\phi_{\epsilon}(\vec{y})\;\frac{\epsilon^{-d}}{ln^{2}\epsilon}\;
\frac{1}{|\;
\vec{x}-\vec{y}\;|^{d}}\; .
\label{3303}
\eeq
Note that the whole functional $I^{\nu =0}_{D}[\phi_{\epsilon}]$ contains,
in addition, the
usual
divergent local terms, plus new logarithmic divergences, which are not
present for $\nu$ not
integer. However, when the constraint Eq.(\ref{150}) is satisfied, the 
usual divergent local terms cancel out, and only the logarithmic
divergences survive. As pointed out before, they
correspond to conformal anomalies, and are to be cancelled out through the
addition of proper counterterms. 

Taking the limit \cite{our2}

\beq
\lim _{\epsilon\rightarrow
0}\; (\epsilon^{\frac{d}{2}} \; ln\;\epsilon)^{-1} \;
\phi_{\epsilon}(\vec{x}) = \phi_{0}(\vec{x}),
\label{3304}
\eeq
in Eq.(\ref{3303}), we find the following boundary two-point function

\beq
\left<{\cal O}_{D}^{\nu=0}(\vec{x})\; {\cal
O}_{D}^{\nu=0}(\vec{y})\right>=
\frac{\Gamma\left(\frac{d}{2}\right)}{2\pi^{\frac{d}{2}}}\;
\frac{1}{|\;
\vec{x}-\vec{y}\;|^{d}}\; .
\label{3305}
\eeq
Then, the operator ${\cal O}_{D}^{\nu=0}$ has conformal dimension
$\frac{d}{2}$, as expected. The classical bulk modes approach the
boundary as
$x_{0}^{\frac{d}{2}}\ln x_{0}$, and we find the logarithmic behavior
that we have mentioned before.

Now, we consider the transformed functional ${\tilde I}^{\nu
=0}_{D}[{\tilde\phi}_{\epsilon}]$. Let us first suppose that the
constraint Eq.(\ref{150}) is not satisfied. Then, after
integration over $\vec{k}$, the leading non-local term in
Eq.(\ref{3302}) reads

\beq
{\tilde I}^{non-local}_{D}[{\tilde\phi}_{\epsilon}] =
-\frac{1}{\left(\frac{d}{2}-2\varrho\; d\right)^{2}}\;\frac{\Gamma\left(
\frac{d}{2}
\right)}{4\pi^{\frac{d}{2}}}\; \int d^{d}x \;
d^{d}y\;{\tilde\phi}_{\epsilon}(\vec{x})
\;{\tilde\phi}_{\epsilon}(\vec{y})\;\frac{\epsilon^{-d}}{ln^{2}\epsilon}\;
\frac{1}{|\;
\vec{x}-\vec{y}\;|^{d}}\; .
\label{3306}
\eeq
In addition, the whole functional ${\tilde
I}^{\varrho\not=\varrho^{\pm}_{D}\; ;\;\nu
=0}_{D}[{\tilde\phi}_{\epsilon}]$ contains the usual divergent local
terms, plus logarithmic divergences. 

Taking the limit

\beq
\lim _{\epsilon\rightarrow
0}\; (\epsilon^{\frac{d}{2}} \; ln\;\epsilon)^{-1} \;
{\tilde\phi}_{\epsilon}(\vec{x}) = {\tilde\phi}_{0}(\vec{x}),
\label{3307}
\eeq
we find the boundary two-point function

\beq
\left<{\tilde {\cal
O}}_{D}^{\varrho\not=\varrho^{\pm}_{D}\; ;\;\nu
=0}(\vec{x})\; {\tilde {\cal
O}}_{D}^{\varrho\not=\varrho^{\pm}_{D}\; ;\;\nu
=0}(\vec{y})\right>=
\frac{1}{\left(\frac{d}{2}-2\varrho\; d\right)^{2}}
\;\frac{\Gamma\left(\frac{d}{2}\right)}{2\pi^{\frac{d}{2}}}\;
\frac{1}{|\;
\vec{x}-\vec{y}\;|^{d}}\; .
\label{3308}
\eeq
Then, the operator ${\tilde {\cal
O}}_{D}^{\varrho\not=\varrho^{\pm}_{D}\; ;\;\nu
=0}$ has conformal dimension $\frac{d}{2}$, and the classical bulk modes 
approach the boundary as $x_{0}^{\frac{d}{2}}\ln x_{0}$.

Finally, we consider the case in which the constraint Eq.(\ref{150}) is
satisfied. In this situation we find that, after integration over
$\vec{k}$, the leading non-local term in Eq.(\ref{3302}) reads

\beq
{\tilde I}^{non-local}_{D}[{\tilde\phi}_{\epsilon}] =
-\frac{\Gamma\left(
\frac{d}{2}
\right)}{4\pi^{\frac{d}{2}}}\; \int d^{d}x \;
d^{d}y\;{\tilde\phi}_{\epsilon}(\vec{x})
\;{\tilde\phi}_{\epsilon}(\vec{y})\;\epsilon^{-d}\;
\frac{1}{|\;
\vec{x}-\vec{y}\;|^{d}}\; .
\label{3309}
\eeq
In addition, the whole functional ${\tilde
I}^{\varrho =\varrho^{\pm}_{D}\; ;\;\nu
=0}_{D}[{\tilde\phi}_{\epsilon}]$
contains logarithmic divergences. Note, also, that unlike
the previous cases, the above leading non-local term does not contain
any logarithmic term in $\epsilon$. Then, instead of
Eq.(\ref{3307}), the limit to be taken is

\beq
\lim _{\epsilon\rightarrow
0}\; \epsilon^{-\frac{d}{2}} \;
{\tilde\phi}_{\epsilon}(\vec{x}) = {\tilde\phi}_{0}(\vec{x}).
\label{3310} 
\eeq
Thus, we find the following boundary two-point function  

\beq
\left<{\tilde {\cal
O}}_{D}^{\varrho =\varrho^{\pm}_{D}\; ;\;\nu
=0}(\vec{x})\; {\tilde {\cal
O}}_{D}^{\varrho =\varrho^{\pm}_{D}\; ;\;\nu
=0}(\vec{y})\right>=
\frac{\Gamma\left(\frac{d}{2}\right)}{2\pi^{\frac{d}{2}}}\;
\frac{1}{|\;
\vec{x}-\vec{y}\;|^{d}}\; .
\label{3311}
\eeq
Then, the operator ${\tilde {\cal O}}_{D}^{\varrho
=\varrho^{\pm}_{D}\; ;\;\nu
=0}$ has conformal dimension 
$\frac{d}{2}$. The classical bulk modes approach the boundary as
$x_{0}^{\frac{d}{2}}$, and this is a novel situation, because we have not
found the usual logarithmic behavior. As anticipated, when the
constraint Eq.(\ref{150}) is satisfied, the Legendre transformation
interpolates between conformal operators of the same conformal dimension
$\frac{d}{2}$, but corresponding to classical bulk modes which
differ in their asymptotic behaviors.

\subsection{Type I Mixed Boundary Condition}

Now we consider the case of Type I mixed boundary conditions. The analysis
is analogous to the one performed in the Dirichlet situation, and in
this case we expect, on considering the conformal dimension
$\Delta_{-}(\varrho)$, to reproduce the constraint Eq.(\ref{152}), instead
of
the one in
Eq.(\ref{150}) corresponding to the Dirichlet boundary condition. Besides,
we also expect to find again the constraint Eq.(\ref{12000}), because,
according to the results in Section 3, it
is common to all boundary conditions.

Let $\psi_{\epsilon}^{I}(\vec{k})$ be the Fourier transform of
$\psi_{\epsilon}^{I}(\vec{x})$. Then, from Eqs.(\ref{723}, \ref{3006},
\ref{3043}, \ref{3044}) we find

\beq 
b(\vec{k})=-\frac{\epsilon^{-\frac{d}{2}}}{K_{\nu}(k\epsilon)}\;
\frac{\psi_{\epsilon}^{I}(\vec{k})}{\frac{d}{2}+\nu
+2\varrho\; d-k\epsilon\frac{K_{\nu
+1}(k\epsilon)}{K_{\nu}(k\epsilon)}}\; , 
\label{3135}
\eeq
and, inserting this into Eq.(\ref{3043}), we write $\phi_{\epsilon}$ in
terms of the boundary data $\psi^{I}_{\epsilon}$ as

\beq
\phi_{\epsilon}(\vec{x})=-\int
d^{d}y\;\psi^{I}_{\epsilon}(\vec{y})\int\frac{d^{d}k}{\left(
2\pi\right)^{d}}\;e^{-i\vec{k}\cdot\left(
\vec{x}-\vec{y}\right)}\;\frac{1}{\frac{d}{2} + \nu +2\varrho\; d-
k\epsilon\;\frac{K_{\nu+1}(k\epsilon)}{K_{\nu}(k\epsilon)}}\; .
\label{3136}
\eeq
Then, the action Eq.(\ref{3041}) can be written as

\beq
I_{M,I}\left[\psi^{I}_{\epsilon}\right]=\frac{1}{2}\int
d^{d}x
\; d^{d}y\;\sqrt{h}\;\psi^{I}_{\epsilon}(\vec{x})
\;\psi^{I}_{\epsilon}(\vec{y})\int\frac{d^{d}k}{\left(
2\pi\right)^{d}}\;e^{-i\vec{k}\cdot\left(
\vec{x}-\vec{y}\right)}\;\frac{1}{\frac{d}{2} + \nu +2\varrho\; d- 
k\epsilon\;\frac{K_{\nu+1}(k\epsilon)}{K_{\nu}(k\epsilon)}}\; .
\label{3137}
\eeq

We also need to compute the corresponding Legendre transformed
functional. This is done by considering Eq.(\ref{9045}) with the
identifications $A_{\epsilon}(\vec{x})\equiv\psi^{I}_{\epsilon}(\vec{x})$,
${\tilde 
A}_{\epsilon}(\vec{x})\equiv{\tilde\psi}^{I}_{\epsilon}(\vec{x})$. Then,
the Legendre transformation is of the form

\beq
{\cal J}_{M,I}\left[\psi^{I}_{\epsilon},{\tilde
\psi}^{I}_{\epsilon}\right]\;=\;  
I_{M,I}\left[\psi^{I}_{\epsilon}\right]\; -\int
d^{d}x\;\sqrt{h}\;\psi^{I}_{\epsilon}(\vec{x})\;
{\tilde\psi}^{I}_{\epsilon}(\vec{x})\; ,
\label{3138}
\eeq
which, after setting $\frac{\partial {\cal
J}_{M,I}}{\partial\psi^{I}_{\epsilon}}=0$, yields

\beq
\psi^{I}_{\epsilon}=\left[\frac{d}{2} + \nu +2\varrho\; d-
k\epsilon\;\frac{K_{\nu+1}(k\epsilon)}{K_{\nu}(k\epsilon)}\right]
\;{\tilde\psi}^{I}_{\epsilon}\; .
\label{3139}
\eeq
Then, the Legendre transform of $I_{M,I}\left[\psi^{I}_{\epsilon}\right]$
is of the form

\ba
{\tilde
I}_{M,I}\left[{\tilde\psi}^{I}_{\epsilon}\right] &=& -\frac{1}{2}\int
d^{d}x
\; d^{d}y\;\sqrt{h}\;{\tilde\psi}^{I}_{\epsilon}(\vec{x})
\;{\tilde\psi}^{I}_{\epsilon}(\vec{y})\nonumber\\
&&\qquad\times\int\frac{d^{d}k}{\left(
2\pi\right)^{d}}\;e^{-i\vec{k}\cdot\left(
\vec{x}-\vec{y}\right)}\;\left[\frac{d}{2} + \nu +2\varrho\; d-
k\epsilon\;\frac{K_{\nu+1}(k\epsilon)}{K_{\nu}(k\epsilon)}\right]. 
\label{3140}
\ea

The functionals Eqs.(\ref{3137}, \ref{3140}) contain the information about
the boundary CFT's corresponding to the Type I mixed boundary-value
problem. Note that, in this case, the generalized AdS/CFT prescription
Eqs.(\ref{9044}, \ref{9044'}) reads

\beq
\exp\left( -I_{M,I}[\psi^{I}_{0}]\right) \equiv
\left<\exp\left(\int
d^{d}x \;
{\cal O}_{M,I}(\vec{x}) \; \psi^{I}_{0}(\vec{x})\right)\right>,
\label{3141}
\eeq
for the original functional, and

\beq
\exp\left( -{\tilde I}_{M,I}[{\tilde\psi}^{I}_{0}]\right) \equiv
\left<\exp\left(\int
d^{d}x \;
{\tilde {\cal
O}}_{M,I}(\vec{x}) \; {\tilde\psi}^{I}_{0}(\vec{x})\right)\right>,
\label{3142} 
\eeq
for the Legendre transformed one. The calculations are analogous to the
ones performed in the Dirichlet case. As expected, the conformal dimension
$\Delta_{-}(\varrho)$ arises precisely when the constraints
Eqs.(\ref{12000}, \ref{152}) are satisfied, and in such situation the
divergent local terms cancel out. In this case, the addition of
counterterms is not required. An interesting feature of this boundary
condition is that the conformal dimension
$\Delta_{-}(\varrho)$ arises from the original functional
Eq.(\ref{3137}), rather than from its corresponding
Legendre transformed functional Eq.(\ref{3140}), which gives rise to
the conformal dimension $\Delta_{+}(\varrho)$. This is a novel
situation, since, as seen before, in the Dirichlet case it is the
transformed functional which gives rise to the conformal dimension
$\Delta_{-}(\varrho)$. As before, there exist particular situations
for which the unitarity
bound $(d-2)/2$ arises. In such cases, the conformal dimension does not
depend
on the effective mass Eq.(\ref{9016'}). Since the calculations are
analogous
to those considered in the Dirichlet case, we just present the main 
results in Appendix B.

\subsection{Type II Mixed Boundary Condition}

The analysis of the Type II mixed situation is analogous to the ones
performed in the Dirichlet and Type I mixed cases. However, it must be
taken into account that, in this case, we have to set $\varrho =0$ in all
calculations, because, unlike the cases considered before, the Type II
mixed boundary conditions are considered only in the minimally coupled
case, as seen in Section 3. Note that now we expect to find the
constraint Eq.(\ref{154}), instead of the ones Eqs.(\ref{150},
\ref{152}) that we have considered so far. Besides, we expect to
reproduce the constraint Eq.(\ref{12000}) in the particular case of
$\varrho =0$.

Let $\psi_{\epsilon}^{I\! I}(\vec{k})$ be the Fourier transform of
$\psi_{\epsilon}^{I\! I}(\vec{x})$. From Eqs.(\ref{728}, \ref{3043},
\ref{3044}) we find

\beq
b(\vec{k})=\frac{\epsilon^{-\frac{d}{2}}}{K_{\nu}(k\epsilon)}\;
\frac{\psi_{\epsilon}^{I\! I}(\vec{k})}
{1-2\lambda\left(\frac{d}{2}+\nu\right)+2\lambda
\;k\epsilon\frac{K_{\nu
+1}(k\epsilon)}{K_{\nu}(k\epsilon)}}\;{\bigg |}_{\varrho =0}\; ,
\label{3150}
\eeq
and using Eq.(\ref{3044}), the above equation allows to write
$\partial_{n}\phi_{\epsilon}$ in terms of the boundary data
$\psi^{I\! I}_{\epsilon}$ as

\beq
\partial_{n}\phi_{\epsilon}(\vec{x})=-\int d^{d}y\;
\psi^{I\! I}_{\epsilon}(\vec{y})\int\frac{d^{d}k}{\left(
2\pi\right)^{d}}\;e^{-i\vec{k}\cdot\left(
\vec{x}-\vec{y}\right)}\;\frac{\frac{d}{2} + \nu
-k\epsilon\;\frac{K_{\nu+1}(k\epsilon)}{K_{\nu}(k\epsilon)}}
{1-2\lambda\left(\frac{d}{2}+
\nu\right)+2\lambda \; k\epsilon\;\frac{K_{\nu+1}(k\epsilon)}
{K_{\nu}(k\epsilon)}}\;{\bigg |}_{\varrho =0}\; .
\label{3151}
\eeq
Thus, the action Eq.(\ref{3042}) reads

\ba
I_{M,I\! I}\left[\psi^{I\! I}_{\epsilon}\right] &=& -\frac{1}{2}\int
d^{d}x
\; d^{d}y\;\sqrt{h}\;\psi^{I\! I}_{\epsilon}(\vec{x})
\;\psi^{I\! I}_{\epsilon}(\vec{y})
\nonumber\\
&& \qquad\qquad\times\int\frac{d^{d}k}{\left(
2\pi\right)^{d}}\;e^{-i\vec{k}\cdot\left(
\vec{x}-\vec{y}\right)}\;\frac{\frac{d}{2} + \nu
-k\epsilon\;\frac{K_{\nu+1}(k\epsilon)}{K_{\nu}(k\epsilon)}}
{1-2\lambda\left(\frac{d}{2}+ 
\nu\right)+2\lambda \; k\epsilon\;\frac{K_{\nu+1}(k\epsilon)}
{K_{\nu}(k\epsilon)}}\;{\bigg |}_{\varrho =0}\; .\nonumber\\
\label{3153}
\ea

Note that, in this case, the Legendre transformation Eq.(\ref{9045}) must
be
performed with the identifications 
$A_{\epsilon}(\vec{x})\equiv\psi^{I\! I}_{\epsilon}(\vec{x})$,
${\tilde
A}_{\epsilon}(\vec{x})\equiv{\tilde\psi}^{I\! I}_{\epsilon}(\vec{x})$.
Then, it reads

\beq
{\cal J}_{M,I\! I}\left[\psi^{I\! I}_{\epsilon},{\tilde
\psi}^{I\! I}_{\epsilon}\right]\;=\;
I_{M,I\! I}\left[\psi^{I\! I}_{\epsilon}\right]\; -\int
d^{d}x\;\sqrt{h}\;\psi^{I\! I}_{\epsilon}(\vec{x})\;
{\tilde\psi}^{I\! I}_{\epsilon}(\vec{x})\; ,
\label{3154}
\eeq
which, after setting $\frac{\partial {\cal
J}_{M,I\! I}}{\partial\psi^{I\! I}_{\epsilon}}=0$, gives rise to

\beq
\psi^{I\! I}_{\epsilon}=-\frac{1-2\lambda\left(\frac{d}{2}+
\nu\right)+2\lambda \; k\epsilon\;\frac{K_{\nu+1}(k\epsilon)}
{K_{\nu}(k\epsilon)}}{\frac{d}{2}+\nu -  
k\epsilon\;\frac{K_{\nu+1}(k\epsilon)}{K_{\nu}(k\epsilon)}}
\;{\tilde\psi}^{I\! I}_{\epsilon}\;{\bigg |}_{\varrho =0}\; .
\label{3155}
\eeq
Then, we find the following Legendre transform of
$I_{M,I\! I}\left[\psi^{I\! I}_{\epsilon}\right]$

\ba
{\tilde I}_{M,I\! I}\left[{\tilde\psi}^{I\! I}_{\epsilon}\right] &=&
\frac{1}{2}\int
d^{d}x
\; d^{d}y\;\sqrt{h}\;{\tilde\psi}^{I\! I}_{\epsilon}(\vec{x})   
\;{\tilde\psi}^{I\! I}_{\epsilon}(\vec{y})
\nonumber\\
&& \qquad\qquad\times\int\frac{d^{d}k}{\left(
2\pi\right)^{d}}\;e^{-i\vec{k}\cdot\left(
\vec{x}-\vec{y}\right)}\;
\frac{1-2\lambda\left(\frac{d}{2}+
\nu\right)+2\lambda \; k\epsilon\;\frac{K_{\nu+1}(k\epsilon)}
{K_{\nu}(k\epsilon)}}
{\frac{d}{2} + \nu
-k\epsilon\;\frac{K_{\nu+1}(k\epsilon)}{K_{\nu}(k\epsilon)}}
\;{\bigg |}_{\varrho =0}\; .\nonumber\\
\label{3156}
\ea

In this case, the generalized AdS/CFT prescription
Eqs.(\ref{9044}, \ref{9044'}) reads

\beq
\exp\left( -I_{M,I\! I}[\psi^{I\! I}_{0}]\right) \equiv
\left<\exp\left(\int  
d^{d}x \;
{\cal O}_{M,I\! I}(\vec{x}) \; \psi^{I\! I}_{0}(\vec{x})\right)\right>,
\label{3157}
\eeq
for the original functional, and

\beq
\exp\left( -{\tilde I}_{M,I\! I}[{\tilde\psi}^{I\! I}_{0}]\right) \equiv
\left<\exp\left(\int
d^{d}x \;
{\tilde {\cal
O}}_{M,I\! I}(\vec{x}) \; {\tilde\psi}^{I\! I}_{0}(\vec{x})\right)\right>, 
\label{3158} 
\eeq
for the Legendre transformed one. The calculations are analogous to the
ones performed in the previous cases, and, as expected, the conformal
dimension $\Delta_{-}(0)$ arises precisely when the constraints
Eqs.(\ref{12000}, \ref{154}) hold (we recall that the constraint
Eq.(\ref{12000}) is to be considered in the particular case of $\varrho
=0$). Also as expected, when the  constraints
Eqs.(\ref{12000}, \ref{154}) are satisfied we find that the divergent
local terms cancel out. As in the former cases, there exist particular
situations for which the unitarity bound $(d-2)/2$ arises. Since the
calculations are analogous to those considered in the previous cases, we
just present the main results in Appendix B.

\section{Conclusions}

In this work, we have revisited the formulation of the scalar field
theory on AdS spaces and in the AdS/CFT correspondence, in both
minimal and non-minimal coupled cases. We have shown that, when quantizing
the scalar field on AdS along the lines of
\cite{freedman}\cite{freedman8}\cite{mezincescu}, there arise some
constraints which cannot be reproduced through the usual AdS/CFT
formalism. In addition, we have found some difficulties
regarding the usual prescription for the Legendre transform,
namely that it leaves a
coefficient to be fixed `by hand', and that it does not work for $\nu 
=0$. In order to remove this
obstacles, we have considered a rather different formulation both in the
bulk and in the AdS/CFT correspondence context.

Regarding the formulation in the bulk, we have argued that the usual
energy which is constructed out of the
stress-energy tensor, namely the metrical energy, is not the natural one
to be considered in the
context of the AdS/CFT correspondence. Then, we have proposed a new
definition of the energy, which makes use of the Noether current
corresponding to time displacements in global coordinates. We have
shown that this new energy, namely the canonical energy, depends on the
boundary conditions, and we have computed it for Dirichlet, Neumann
and mixed boundary-value problems, in both minimally
and non-minimally coupled cases. By requiring this energy to be 
conserved, positive and finite, we have found that the
Breitenlohner-Freedman bound still holds, that regular and irregular modes
propagate in the bulk as it happens when the quantization makes
use of the metrical energy, and that there arise new constraints for
which the irregular modes propagate.

Then, we have shown that this new formalism gives rise, in a
natural way, to a generalized AdS/CFT prescription in which the source
which couples to the boundary conformal operator depends on the selected
boundary condition, thus involving a situation which is much more general
than the usual one, where only Dirichlet boundary conditions are
considered. We have also analyzed a generalized prescription for
the Legendre transform that, in fact, makes use of the standard form of
the Legendre transformation. This means, in particular, that we transform
the whole on-shell action, rather than only the leading non-local term, as
in \cite{witten2}. In this generalized prescription, there is no
coefficient to be fixed `by hand', and the case $\nu =0$ can be analyzed
in a natural way.

We have shown that this generalized AdS/CFT prescription gives rise, in a
natural way, to the constraints for which irregular modes propagate in the
bulk when the canonical energy is taken into account, thus providing
strong evidence in support of our formalism. Another
interesting result that we have shown is that, for such constraints, the
divergent local terms of the on-shell action cancel out, and the Legendre
transformation interpolates between different conformal dimensions, as
expected. In other words, the divergent local terms contain information
about the transformed generating functional, and then they have to be
taken into account when performing the Legendre transformation. In the
particular situation in which the divergent local terms cancel out and the
conformal dimension $\Delta_{-}(\varrho)$ arises, the addition of
counterterms to the on-shell action is not required. 

Finally, we have also shown that there exists one particular case which
has no bulk counterpart, namely the one in which the conformal dimension
reaches the unitarity bound and becomes independent of the effective
mass. This puzzling phenomenon deserves further studies.

\section{Acknowledgements}

P.M. acknowledges financial support by FAPESP grant 01/05770-1. V.O.R. is
partially supported by CNPq and PRONEX under contract CNPq
66.2002/1998-99.

\appendix

\section{Hypergeometric Functions ${\bf _{2}F_{1}}$}

\beq
_{2}F_{1}(a,b\; ;c\; ;x)=\frac{\Gamma(c)}{\Gamma(a)\Gamma(b)}\;\sum_{n\geq
0}
\frac{\Gamma(a+n)\Gamma(b+n)}{\Gamma(c+n)}\;\frac{x^n}{n!}\; .
\label{922}
\eeq
\\
i) $\;$ For $c-a-b\not\in {\bf Z}$

\ba
_{2}F_{1}(a,b\; ;c\; ;x)&=&\frac{\Gamma(c)\Gamma(c-a-b)}
{\Gamma(c-a)\Gamma(c-b)}\; {_{2}F_{1}}
(a,b\; ;a+b-c+1\; ;1-x)\nonumber\\
&+&(1-x)^{c-a-b}\;\frac{\Gamma(c)\Gamma(a+b-c)}
{\Gamma(a)\Gamma(b)}\nonumber\\ &\times &
{_{2}F_{1}}(c-a,c-b\; ;c-a-b+1\; ;1-x)\; .
\label{923}
\ea
\\
ii) $\;$ For $c-a-b=0$

\ba
_{2}F_{1}(a,b\; ;c\; ;x)&=&-\; \frac{\Gamma(a+b)}
{[\Gamma(a)]^{2}[\Gamma(b)]^{2}}\;\sum_{n\geq 0}
\frac{\Gamma (a+n)\Gamma (b+n)}{[\Gamma(n+1)]^{2}}\nonumber\\
&\times& \left[ ln(1-x)+u_{n}(a,b,0)\right](1-x)^{n}\; ,\nonumber\\
\label{924}
\ea
where

\beq
u_{n}(a,b,m)=\beta(n+a+m)+\beta(n+b+m)-\beta(n+1)-\beta(n+1+m)\; ,
\label{924'}
\eeq
\beq
\beta(z)=\frac{d}{dz}\; ln \;\Gamma (z)\; ,
\label{925}
\eeq

\beq
\beta(1)=-\gamma \qquad\qquad \beta(n)=-\gamma +
\sum_{p=1}^{n-1}\frac{1}{p}\qquad (n=2,3,\cdots),
\label{926}
\eeq
and $\gamma$ is the Euler constant.   
\\
\\
\\
\\
\\
iii) $\;$ For $c-a-b=m\quad (m=1,2,\cdots )$

\ba
_{2}F_{1}(a,b\; ;c\; ;x)&=&
\frac{\Gamma(m)\Gamma(a+b+m)\Gamma(1-m)}{\Gamma(a)\Gamma(b)
\Gamma(a+m)\Gamma(b+m)}\;
\sum_{n=0}^{m-1}\frac{\Gamma(n+a)\Gamma(n+b)}{\Gamma(n+1)\Gamma(n+1-m)}\;
(1-x)^{n}\nonumber\\
&-&\; (-1)^{m}\frac{\Gamma(a+b+m)}
{\Gamma(a)\Gamma(b)\Gamma(a+m)\Gamma(b+m)}\;\sum_{n\geq 0}
\frac{\Gamma (n+a+m)\Gamma (n+b+m)}{\Gamma(n+1)\Gamma(n+1+m)}\nonumber\\
&\times& \left[ ln(1-x)+u_{n}(a,b,m)\right](1-x)^{n+m}\; .
\label{927}
\ea
\\
iv) $\;$ For $c-a-b=-m\quad (m=1,2,\cdots )$

\ba
_{2}F_{1}(a,b\; ;c\; ;x)&=&
\frac{\Gamma(m)\Gamma(a+b-m)\Gamma(1-m)}{\Gamma(a)\Gamma(b)
\Gamma(a-m)\Gamma(b-m)}\nonumber\\ &\times &
\sum_{n=0}^{m-1}\frac{\Gamma(n+a-m)\Gamma(n+b-m)}{\Gamma(n+1)\Gamma(n+1-m)}\;
(1-x)^{n-m}\nonumber\\
&-&\; (-1)^{m}\frac{\Gamma(a+b-m)}
{\Gamma(a)\Gamma(b)\Gamma(a-m)\Gamma(b-m)}\;\sum_{n\geq 0}
\frac{\Gamma (n+a)\Gamma (n+b)}{\Gamma(n+1)\Gamma(n+1+m)}\nonumber\\
&\times& \left[ ln(1-x)+v_{n}(a,b,m)\right](1-x)^{n}\; ,
\label{928}
\ea
where
\beq
v_{n}(a,b,m)=\beta(n+a)+\beta(n+b)-\beta(n+1)-\beta(n+1+m)\; .
\label{929}
\eeq

\section{Results Obtained from the Generalized AdS/CFT Prescription}

In this appendix, we summarize the main results which are obtained through
the generalized AdS/CFT prescription Eqs.(\ref{9044}-\ref{9044'}), for
Dirichlet and Types I and II mixed boundary
conditions. Let us first consider the case when $\nu$ is not an 
integer. We 
begin by analyzing a generic result, and then we will specialize to
each specific situation. In general, we find generating functionals for
the boundary CFT's which are of the generic form

\ba
{\cal F}[f_{\epsilon}]= -\;{\cal C}\int d^{d}x \;
d^{d}y\; f_{\epsilon}(\vec{x})
\; f_{\epsilon}(\vec{y})
\;\frac{\epsilon^{-2\Delta_{1}}}{|\;
\vec{x}-\vec{y}\;|^{2\Delta_{2}}}\; ,
\label{8000}
\ea
where ${\cal C}$ is a coefficient, $f_{\epsilon}$ is the classical
background
which encodes the information about the boundary CFT, $\Delta_{1}$ is
to be related to the asymptotic behavior of $f_{\epsilon}$ and
$\Delta_{2}$
corresponds to the conformal dimension of the boundary operator. Taking
the limit

\beq
\lim _{\epsilon\rightarrow
0}\epsilon^{-\Delta_{1}}\; f_{\epsilon}(\vec{x})
= f_{0}(\vec{x}),
\label{8001}
\eeq
and using the AdS/CFT prescription

\beq
\exp\left( -{\cal F}\right) \equiv \left<\exp\left(\int d^{d}x \; 
{\cal O}_{f}(\vec{x}) \; f_{0}(\vec{x})\right)\right>,
\label{8002}
\eeq
we find the boundary two-point function

\beq
\left<{\cal O}_{f}(\vec{x})\; {\cal
O}_{f}(\vec{y})\right> = 2\;{\cal C}
\;\frac{1}{|\;
\vec{x}-\vec{y}\;|^{2\Delta_{2}}}\; .
\label{8003}
\eeq
Then, the field $f_{\epsilon}$ has the asymptotic behavior
$\epsilon^{\Delta_{1}}$, and the operator ${\cal O}_{f}$ has conformal
dimension $\Delta_{2}$.

Now we are ready to present the main results obtained for the specific
cases of Dirichlet and Types I and II mixed boundary conditions. We
consider each case separately. Recall that, in particular, the unitarity
bound $(d-2)/2$ arises only for $d\geq 3$.

\subsection{Dirichlet}

\bigskip
\bigskip

\begin{center}
\begin{tabular}{c|c|c|c|c}
& $\Delta_{1}$  & $\Delta_{2}$ & ${\cal C}$ & Divergences
\\ \hline
& & & & \\  
$f_{\epsilon}=\phi_{\epsilon}$  & $\Delta_{-}(\varrho)$
  & $\Delta_{+}(\varrho)$  & $\frac{\nu (\varrho)}{\pi^{\frac{d}{2}}}\;
\frac{\Gamma(\Delta_{+}(\varrho))}{\Gamma(\nu (\varrho))}$
& YES \\ 
& & & & \\
$f_{\epsilon}={\tilde \phi}_{\epsilon}$  & $\Delta_{-}(\varrho)$ 
& $\Delta_{+}(\varrho)$ & 
$\frac{\nu (\varrho)}{\pi^{\frac{d}{2}}}\;
\frac{1}{\left[\Delta_{-}(\varrho)-2\varrho\; d\right]^{2}}
\;\frac{\Gamma(\Delta_{+}(\varrho))}{\Gamma(\nu (\varrho))}$
 & YES \\
\end{tabular}
\end{center}
\centerline{{\bf Table 1:} Values of $\Delta_1, \Delta_2$ and $\cal{C}$
and presence of divergences when Eq.(\ref{150}) is not}

\hspace{1.5cm} satisfied.

\bigskip
\bigskip
\bigskip

\begin{center}
\begin{tabular}{c|c|c|c|c}
& $\Delta_{1}$  & $\Delta_{2}$ & ${\cal C}$ & Divergences
\\ \hline
& & & & \\  
$f_{\epsilon}=\phi_{\epsilon}$  & $\Delta_{-}(\varrho)$
  & $\Delta_{+}(\varrho)$  & $\frac{\nu (\varrho)}{\pi^{\frac{d}{2}}}\;
\frac{\Gamma(\Delta_{+}(\varrho))}{\Gamma(\nu (\varrho))}$
& NO \\
& & & & \\
$f_{\epsilon}={\tilde \phi}_{\epsilon}$  & $\Delta_{+}(\varrho)$
& $\Delta_{-}(\varrho)$ &
$\frac{1}{4\pi^{\frac{d}{2}}}
\;\frac{\Gamma(\Delta_{-}(\varrho))}{\Gamma(1-\nu(\varrho))} $
 & NO \\
\end{tabular}
\end{center}
\centerline{{\bf Table 2:} Values of $\Delta_1, \Delta_2$ and $\cal{C}$
and presence of divergences when Eqs.(\ref{12000}, \ref{150}) are}

\hspace{1.3cm} satisfied.

\bigskip
\bigskip
\bigskip

\begin{center}
\begin{tabular}{c|c|c|c|c}
& $\Delta_{1}$  & $\Delta_{2}$ & ${\cal C}$ & Divergences
\\ \hline
& & & & \\
$f_{\epsilon}=\phi_{\epsilon}$  & $\Delta_{-}(\varrho)$
  & $\Delta_{+}(\varrho)$  & $\frac{\nu (\varrho)}{\pi^{\frac{d}{2}}}\;
\frac{\Gamma(\Delta_{+}(\varrho))}{\Gamma(\nu (\varrho))}$
& YES \\
& & & & \\
$f_{\epsilon}={\tilde \phi}_{\epsilon}$  & $\frac{d+2}{2}$    
& $\frac{d-2}{2}$ &
$\left(\nu(\varrho)-1\right)
\frac{\Gamma(\frac{d-2}{2})}{4\pi^{\frac{d}{2}}}$
 & NO \\
\end{tabular} 
\end{center}
\centerline{{\bf Table 3:} Values of $\Delta_1, \Delta_2$ and $\cal{C}$
and presence of divergences when Eq.(\ref{150}) is}

\hspace{1.9cm} satisfied, but Eq.(\ref{12000}) is not.

\subsection{Type I Mixed}

\bigskip
\bigskip

\begin{center}
\begin{tabular}{c|c|c|c|c}
& $\Delta_{1}$  & $\Delta_{2}$ & ${\cal C}$ & Divergences
\\ \hline
& & & & \\
$f_{\epsilon}=\psi^{I}_{\epsilon}$  & $\Delta_{-}(\varrho)$
  & $\Delta_{+}(\varrho)$  & 
$\frac{\nu (\varrho)}{\pi^{\frac{d}{2}}}\;
\frac{1}{\left[\Delta_{-}(\varrho)+2\varrho\; d\right]^{2}}
\;\frac{\Gamma(\Delta_{+}(\varrho))}{\Gamma(\nu (\varrho))}$
& YES \\
& & & & \\
$f_{\epsilon}={\tilde \psi}^{I}_{\epsilon}$ 
 & $\Delta_{-}(\varrho)$
& $\Delta_{+}(\varrho)$ &
$\frac{\nu (\varrho)}{\pi^{\frac{d}{2}}}\;
\frac{\Gamma(\Delta_{+}(\varrho))}{\Gamma(\nu (\varrho))}$
 & YES \\
\end{tabular}
\end{center}
\centerline{{\bf Table 4:} Values of $\Delta_1, \Delta_2$ and $\cal{C}$
and presence of divergences when Eq.(\ref{152}) is not}

\hspace{1.5cm} satisfied.

\bigskip 
\bigskip  
\bigskip

\begin{center}
\begin{tabular}{c|c|c|c|c}
& $\Delta_{1}$  & $\Delta_{2}$ & ${\cal C}$ & Divergences
\\ \hline
& & & & \\
$f_{\epsilon}=\psi^{I}_{\epsilon}$  & $\Delta_{+}(\varrho)$
  & $\Delta_{-}(\varrho)$
  & $\frac{1}{4\pi^{\frac{d}{2}}}
\;\frac{\Gamma(\Delta_{-}(\varrho))}{\Gamma(1-\nu(\varrho))}$
& NO \\
& & & & \\
$f_{\epsilon}=
{\tilde \psi}^{I}_{\epsilon}$  & $\Delta_{-}(\varrho)$
& $\Delta_{+}(\varrho)$ &
$\frac{\nu (\varrho)}{\pi^{\frac{d}{2}}}\;
\frac{\Gamma(\Delta_{+}(\varrho))}{\Gamma(\nu (\varrho))} $
 & NO \\
\end{tabular}
\end{center}
\centerline{{\bf Table 5:} Values of $\Delta_1, \Delta_2$ and $\cal{C}$
and presence of divergences when Eqs.(\ref{12000}, \ref{152}) are}

\hspace{1.3cm} satisfied.

\bigskip
\bigskip
\bigskip

\begin{center}
\begin{tabular}{c|c|c|c|c}
& $\Delta_{1}$  & $\Delta_{2}$ & ${\cal C}$ & Divergences
\\ \hline
& & & & \\
$f_{\epsilon}=\psi^{I}_{\epsilon}$  & $\frac{d+2}{2}$
  & $\frac{d-2}{2}$  & 
$\left(\nu(\varrho)-1\right)
\frac{\Gamma(\frac{d-2}{2})}{4\pi^{\frac{d}{2}}}$
& NO \\
& & & & \\
$f_{\epsilon}={\tilde \psi}^{I}_{\epsilon}$  &
$\Delta_{-}(\varrho)$
& $\Delta_{+}(\varrho)$ &
$\frac{\nu (\varrho)}{\pi^{\frac{d}{2}}}\;
\frac{\Gamma(\Delta_{+}(\varrho))}{\Gamma(\nu (\varrho))}$
 & YES \\  
\end{tabular}
\end{center}
\centerline{{\bf Table 6:} Values of $\Delta_1, \Delta_2$ and $\cal{C}$
and presence of divergences when Eq.(\ref{152}) is}

\hspace{1.9cm} satisfied, but Eq.(\ref{12000}) is not.

\subsection{Type II Mixed}

In this case, we also need to consider the additional constraint

\beq
m=0\; .
\label{6278}
\eeq

\bigskip

\begin{center}
\begin{tabular}{c|c|c|c|c}
& $\Delta_{1}$  & $\Delta_{2}$ & ${\cal C}$ & Divergences
\\ \hline
& & & & \\
$f_{\epsilon}=\psi^{I\! I}_{\epsilon}$  & $\Delta_{-}(0)$
  & $\Delta_{+}(0)$  & 
$\frac{\nu (0)}{\pi^{\frac{d}{2}}}\;
\frac{1}{\left[1-2\lambda\;\Delta_{-}(0)\right]^{2}}
\;\frac{\Gamma(\Delta_{+}(0))}{\Gamma(\nu (0))}$
& YES \\
& & & & \\  
$f_{\epsilon}={\tilde \psi}^{I\! I}_{\epsilon}$ 
 & $\Delta_{-}(0)$
& $\Delta_{+}(0)$ &
$\frac{\nu (0)}{\pi^{\frac{d}{2}}}\;   
\frac{1}{\left[\Delta_{-}(0)\right]^{2}}   
\;\frac{\Gamma(\Delta_{+}(0))}{\Gamma(\nu (0))}$
 & YES \\
\end{tabular} 
\end{center}
\centerline{{\bf Table 7:} Values of $\Delta_1, \Delta_2$ and $\cal{C}$
and presence of divergences when Eqs.(\ref{154}, \ref{6278}) are}

\hspace{1.3cm} not satisfied.

\bigskip
\bigskip

\begin{center}
\begin{tabular}{c|c|c|c|c}
& $\Delta_{1}$  & $\Delta_{2}$ & ${\cal C}$ & Divergences
\\ \hline
& & & & \\
$f_{\epsilon}=\psi^{I\! I}_{\epsilon}$  & $\Delta_{+}(0)$
  & $\Delta_{-}(0)$  & 
$\frac{1}{4\pi^{\frac{d}{2}}}\;
\left[\Delta_{-}(0)\right]^{2}
\;\frac{\Gamma(\Delta_{-}(0))}{\Gamma(1-\nu(0))}$
& NO \\
& & & & \\
$f_{\epsilon}={\tilde \psi}^{I\! I}_{\epsilon}$
  & $\Delta_{-}(0)$
& $\Delta_{+}(0)$ &
$\frac{\nu (0)}{\pi^{\frac{d}{2}}}\;
\frac{1}{\left[\Delta_{-}(0)\right]^{2}}
\;\frac{\Gamma(\Delta_{+}(0))}{\Gamma(\nu (0))}$
 & NO \\
\end{tabular}
\end{center}
\centerline{{\bf Table 8:} Values of $\Delta_1, \Delta_2$ and $\cal{C}$
and presence of divergences when Eqs.(\ref{12000}, \ref{154}) are}

\hspace{1.3cm} satisfied.

\bigskip
\bigskip

\begin{center} 
\begin{tabular}{c|c|c|c|c}
& $\Delta_{1}$  & $\Delta_{2}$ & ${\cal C}$ & Divergences
\\ \hline
& & & & \\
$f_{\epsilon}=\psi^{I\! I}_{\epsilon}$  & $\frac{d+2}{2}$
  & $\frac{d-2}{2}$  & 
$\left(\nu(0)-1\right)\;\left[\Delta_{-}(0)\right]^{2}   
\;\frac{\Gamma(\frac{d-2}{2})}{4\pi^{\frac{d}{2}}}$
& NO \\
& & & & \\
$f_{\epsilon}={\tilde \psi}^{I\! I}_{\epsilon}$ 
 & $\Delta_{-}(0)$
& $\Delta_{+}(0)$ &
$\frac{\nu (0)}{\pi^{\frac{d}{2}}}\;
\frac{1}{\left[\Delta_{-}(0)\right]^{2}}
\;\frac{\Gamma(\Delta_{+}(0))}{\Gamma(\nu (0))}$
 & YES \\  
\end{tabular}
\end{center}
\centerline{{\bf Table 9:} Values of $\Delta_1, \Delta_2$ and $\cal{C}$
and presence of divergences when Eq.(\ref{154}) is}

\hspace{1.9cm} satisfied, but Eq.(\ref{12000}) is not.
 
\bigskip
\bigskip
\bigskip

\begin{center}
\begin{tabular}{c|c|c|c|c}
& $\Delta_{1}$  & $\Delta_{2}$ & ${\cal C}$ & Divergences
\\ \hline
& & & & \\
$f_{\epsilon}=\psi^{I\! I}_{\epsilon}$  & $\Delta_{-}(0)\;(=0)$
  & $\Delta_{+}(0)\;(=d)$  &
$\frac{d}{2\pi^{\frac{d}{2}}}\;
\;\frac{\Gamma(d)}{\Gamma(\frac{d}{2})}$
& YES \\ 
& & & & \\
$f_{\epsilon}={\tilde \psi}^{I\! I}_{\epsilon}$
 & $\frac{d+2}{2}$
& $\frac{d-2}{2}$ &
$\frac{\Gamma(\frac{d}{2})}{4\pi^{\frac{d}{2}}}$
 & NO \\
\end{tabular}
\end{center}
\centerline{{\bf Table 10:} Values of $\Delta_1, \Delta_2$ and $\cal{C}$
and presence of divergences when Eq.(\ref{6278}) is}

\hspace{1.9cm} satisfied.
 
\bigskip
\bigskip
\bigskip
Now we consider the case of $\nu$ integer. For $\nu$ integer but non-zero,
it can be shown, following a procedure analogous to the one in
\cite{our2}, that, after integrating over the momentum, the leading
non-local terms are equal to the
ones that we have found in the case of $\nu$ not integer, thus
giving rise to the same boundary two-point functions. Besides, in
addition to the usual divergent local terms, there arise logarithmic
divergences, which correspond to conformal anomalies.

To finish, we present the results corresponding to the case $\nu
=0$. As in the
case of $\nu$ not integer, we begin by analyzing a generic result, and
then we will specialize to each specific situation. For $\nu =0$, in
general we find two different kinds of leading non-local terms. They have
the generic form

\ba
{\cal F}_{1}[f_{\epsilon}]= -\;{\cal C}\int d^{d}x \;
d^{d}y\; f_{\epsilon}(\vec{x})
\; f_{\epsilon}(\vec{y})
\;\frac{\epsilon^{-d}}{ln^{2}\epsilon}\;\frac{1}{|\;
\vec{x}-\vec{y}\;|^{d}}\; ,
\label{7001}
\ea
and

\ba
{\cal F}_{2}[f_{\epsilon}]= -\;{\cal C}\int d^{d}x \;
d^{d}y\; f_{\epsilon}(\vec{x})
\; f_{\epsilon}(\vec{y})
\;\frac{\epsilon^{-d}}{|\;
\vec{x}-\vec{y}\;|^{d}}\; .
\label{7000}
\ea
In the first one, we take the limit

\beq
\lim _{\epsilon\rightarrow 
0}\; (\epsilon^{\frac{d}{2}} \; ln\;\epsilon)^{-1} \;
f_{\epsilon}(\vec{x}) = f_{0}(\vec{x}),
\label{7002}
\eeq
which corresponds to the asymptotic behavior $x_{0}^{\frac{d}{2}}\ln
x_{0}$. In the second one, the limit to be taken is

\beq
\lim _{\epsilon\rightarrow
0}\; \epsilon^{-\frac{d}{2}} \;
f_{\epsilon}(\vec{x}) = f_{0}(\vec{x}),
\label{7003}
\eeq
and this corresponds to the asymptotic behavior
$x_{0}^{\frac{d}{2}}$. In both cases, we find the following boundary
two-point function

\beq
\left<{\cal O}_{f}(\vec{x})\; {\cal
O}_{f}(\vec{y})\right> = 2\;{\cal C}
\;\frac{1}{|\;
\vec{x}-\vec{y}\;|^{d}}\; ,
\label{7004}
\eeq
which corresponds to the conformal dimension $\frac{d}{2}$.

Now we are ready to present the main results obtained for the specific
cases of Dirichlet and Types I and II mixed boundary conditions. We
consider each case separately.
   
\subsection{Dirichlet}

\bigskip
\bigskip

\begin{center}
\begin{tabular}{c|c|c}
& ${\cal F}$  & ${\cal C}$
\\ \hline
& & \\
$f_{\epsilon}=\phi_{\epsilon}$  & 
${\cal F}_{1}$
 & 
$\frac{\Gamma\left(
\frac{d}{2}\right)}{4\pi^{\frac{d}{2}}}$
  \\
& & \\
$f_{\epsilon}={\tilde \phi}_{\epsilon}$  &
${\cal F}_{1}$
 &
$\frac{1}{\left(\frac{d}{2}-2\varrho\; d\right)^{2}}
\;\frac{\Gamma\left(
\frac{d}{2}\right)}{4\pi^{\frac{d}{2}}}$
 \\
\end{tabular}
\end{center}
\centerline{{\bf Table 11:} Leading non-local terms and values of
$\cal{C}$ when Eq.(\ref{150}) is not satisfied.}   

\bigskip
\bigskip
\bigskip

\begin{center}
\begin{tabular}{c|c|c}
& ${\cal F}$  & ${\cal C}$
\\ \hline
& & \\
$f_{\epsilon}=\phi_{\epsilon}$  &
${\cal F}_{1}$
 &
$\frac{\Gamma\left(
\frac{d}{2}\right)}{4\pi^{\frac{d}{2}}}$
  \\
& & \\
$f_{\epsilon}={\tilde \phi}_{\epsilon}$  &
${\cal F}_{2}$ 
 &
$\frac{\Gamma\left(
\frac{d}{2}\right)}{4\pi^{\frac{d}{2}}}$
 \\
\end{tabular}
\end{center}
\centerline{{\bf Table 12:} Leading non-local terms and values of
$\cal{C}$ when Eq.(\ref{150}) is satisfied.}

\subsection{Type I Mixed}

\bigskip
\bigskip

\begin{center}
\begin{tabular}{c|c|c}
& ${\cal F}$  & ${\cal C}$
\\ \hline
& & \\
$f_{\epsilon}=\psi^{I}_{\epsilon}$  &
${\cal F}_{1}$
 &
$\frac{1}{\left(\frac{d}{2}+2\varrho\; d\right)^{2}}
\;\frac{\Gamma\left(
\frac{d}{2}\right)}{4\pi^{\frac{d}{2}}}$
  \\
& & \\
$f_{\epsilon}={\tilde \psi}^{I}_{\epsilon}$  &
${\cal F}_{1}$
 &
$\frac{\Gamma\left(
\frac{d}{2}\right)}{4\pi^{\frac{d}{2}}}$
 \\
\end{tabular}
\end{center}
\centerline{{\bf Table 13:} Leading non-local terms and values of
$\cal{C}$ when Eq.(\ref{152}) is not satisfied.}
  
\bigskip
\bigskip

\begin{center}
\begin{tabular}{c|c|c}
& ${\cal F}$  & ${\cal C}$
\\ \hline
& & \\
$f_{\epsilon}=\psi^{I}_{\epsilon}$  &
${\cal F}_{2}$
 &
$\frac{\Gamma\left(
\frac{d}{2}\right)}{4\pi^{\frac{d}{2}}}$
  \\  
& & \\
$f_{\epsilon}={\tilde \psi}^{I}_{\epsilon}$  &
${\cal F}_{1}$
 &
$\frac{\Gamma\left(
\frac{d}{2}\right)}{4\pi^{\frac{d}{2}}}$
 \\
\end{tabular} 
\end{center}
\centerline{{\bf Table 14:} Leading non-local terms and values of
$\cal{C}$ when Eq.(\ref{152}) is satisfied.}  

\bigskip
\bigskip

\subsection{Type II Mixed}

\bigskip
\bigskip

\begin{center}
\begin{tabular}{c|c|c}
& ${\cal F}$  & ${\cal C}$
\\ \hline
& & \\
$f_{\epsilon}=\psi^{I\! I}_{\epsilon}$  &
${\cal F}_{1}$
 &
$\frac{1}{\left(1-\lambda\; d\right)^{2}}
\;\frac{\Gamma\left(
\frac{d}{2}\right)}{4\pi^{\frac{d}{2}}}$
  \\
& & \\
$f_{\epsilon}={\tilde \psi}^{I\! I}_{\epsilon}$  &
${\cal F}_{1}$
 &
$\left(\frac{2}{d}\right)^{2}\;\frac{\Gamma\left(
\frac{d}{2}\right)}{4\pi^{\frac{d}{2}}}$
 \\
\end{tabular}
\end{center}
\centerline{{\bf Table 15:} Leading non-local terms and values of
$\cal{C}$ when Eq.(\ref{154}) is not satisfied.}

\bigskip
\bigskip
\bigskip

\begin{center}
\begin{tabular}{c|c|c}
& ${\cal F}$  & ${\cal C}$
\\ \hline
& & \\
$f_{\epsilon}=\psi^{I\! I}_{\epsilon}$  &
${\cal F}_{2}$
 &
$\left(\frac{d}{2}\right)^{2}\;\frac{\Gamma\left(
\frac{d}{2}\right)}{4\pi^{\frac{d}{2}}}$
  \\  
& & \\
$f_{\epsilon}={\tilde \psi}^{I\! I}_{\epsilon}$  &
${\cal F}_{1}$
 &
$\left(\frac{2}{d}\right)^{2}\;\frac{\Gamma\left(
\frac{d}{2}\right)}{4\pi^{\frac{d}{2}}}$
 \\
\end{tabular} 
\end{center}
\centerline{{\bf Table 16:} Leading non-local terms and values of
$\cal{C}$ when Eq.(\ref{154}) is satisfied.}

\end{document}